\documentclass[preprint]{elsarticle}
\usepackage[textwidth=18cm, textheight=24cm]{geometry}

\usepackage[utf8]{inputenc}
\usepackage[english]{babel}
\usepackage[T1]{fontenc}
\usepackage[hidelinks]{hyperref}
\usepackage{amsmath}
\usepackage{amssymb}
\usepackage{subcaption}
\usepackage{bm}
\usepackage{siunitx}
\usepackage{setspace}
\usepackage{cancel}
\usepackage{nicefrac}

\hypersetup{pdfauthor=author}

\def\etal.{et\penalty50\ al.}

\makeatletter
\def\ps@pprintTitle{%
  \let\@oddhead\@empty
  \let\@evenhead\@empty
  \let\@oddfoot\@empty
  \let\@evenfoot\@oddfoot
}
\makeatother

\begin{document}

\begin{frontmatter}
\title{Spectral adjoint-based assimilation of sparse data in unsteady simulations of turbulent flows}

\author[ifd,empa]{Justin Plogmann\corref{corauthor}}
\cortext[corauthor]{Corresponding author: \url{justin.plogmann@empa.ch}}

\author[ifd]{Oliver Brenner}

\author[ifd]{Patrick Jenny}

\affiliation[ifd]{%
    organization={Institute of Fluid Dynamics, ETH Zurich},%
    addressline={Sonneggstrasse~3},%
    city={Zürich},%
    postcode={CH-8092},%
    country={Switzerland}%
}

\affiliation[empa]{%
    organization={Chemical Energy Carriers and Vehicle Systems Laboratory, Swiss Federal Laboratories for Materials Science and Technology (Empa)},%
    addressline={Überlandstrasse~129},%
    city={Dübendorf},%
    postcode={CH-8600},%
    country={Switzerland}%
}

\begin{abstract}
The unsteady Reynolds-averaged Navier--Stokes (URANS) equations provide a computationally efficient tool to simulate unsteady turbulent flows for a wide range of applications. To account for the errors introduced by the turbulence closure model, recent works have adopted data assimilation (DA) to enhance their predictive capabilities. Recognizing the challenges posed by the computational cost of four-dimensional variational (4DVar) DA for unsteady flows, we propose a three-dimensional (3DVar) DA framework that incorporates a time-discrete Fourier transform of the URANS equations, facilitating the use of the stationary discrete adjoint method in Fourier space.

Central to our methodology is the introduction of a corrective, divergence-free, and unsteady forcing term, derived from a Fourier series expansion, into the URANS equations. This term aims at mitigating discrepancies in the modeled divergence of Reynolds stresses, allowing for the tuning of stationary parameters across different Fourier modes. While designed to accommodate multiple modes in general, the basic capabilities of our framework are demonstrated for a setup that is truncated after the first Fourier mode.

Our implementation is built upon an extended version of the coupled URANS solver in \textit{OpenFOAM}, enhanced to compute adjoint variables and gradients. This design choice ensures straightforward applicability to various flow setups and solvers, eliminating the need for specialized harmonic solvers. A gradient-based optimizer is employed to minimize discrepancies between simulated results and sparse velocity reference data.

The effectiveness of our approach is demonstrated through its application to flow around a two-dimensional circular cylinder at a Reynolds number of 3900. Our results highlight the method's ability to reconstruct mean flow accurately and improve the vortex shedding frequency (Strouhal number) through the assimilation of zeroth mode data. Additionally, the assimilation of first mode data further enhances the simulation's capability to capture low-frequency dynamics of the flow, and finally, it runs efficiently by leveraging a coarse mesh.

\end{abstract}


\end{frontmatter}


\section{Introduction}
\label{sec:intro}

Efficient and accurate analysis of fluid flows is essential in many engineering applications. Such flows are often turbulent since they occur at high Reynolds numbers and involve many length and time scales. The largest scale is often linked to the size of the flow problem. It can therefore be the size of a planet in meteorological applications or the length of an aircraft. These large scales then feed energy into smaller scales until they finally dissipate at the very smallest scales. Resolving the large range of spatial and temporal scales involved in turbulent flows with direct numerical simulations (DNS) or partially resolve the scales by conducting large eddy simulations (LES) is prohibitively expensive and only feasible at low Reynolds numbers~\cite{Pope2000}.

Therefore, the (unsteady) Reynolds-averaged Navier--Stokes ((U)RANS) equations continue to be the workhorse approach for industrial flow problems. This approach models all turbulent fluctuations and only provides (ensemble) averaged results, which makes it very affordable. While URANS is able to resolve the coherent vortex structures, RANS aims to find a steady-state solution and therefore lacks the ability of predicting the time-dependent deterministic vortex shedding~\cite{Frohlich2008}. But also for mean flow empirical evidence supports the superiority of URANS~\cite{Durbin1995}.

(U)RANS models, however, often rely on the Boussinesq hypothesis, which introduces the eddy viscosity and the associated model-form errors~\cite{xiao19,XIAO17}. Duraisamy~\etal.~\cite{duraisamy19} reviews the more recent developments that focus on data driven models or machine learning to close the RANS equations. Physics-informed machine learning (PIML), as demonstrated in Karniadakis~\etal.~\cite{karniadakis21}, aims to incorporate principles of physics into machine learning techniques, including neural networks and methods based on kernels. Wang~\etal.~\cite{wang17} applied a PIML strategy to adjust the modeled Reynolds stresses within RANS equations. Raissi~\etal.~\cite{raissi18} introduced Physics-informed neural networks (PINN) that are trained to solve supervised learning tasks while respecting any given law of physics described by general nonlinear partial differential equations. Sliwinski and Rigas~\cite{sliwinski_mean_2023} performed PINN-based reconstruction of the mean flow around a circular cylinder from sparse time-averaged velocity data to infer unknown closure quantities.

Data assimilation (DA) represents a technique that integrates observational data into simulation models to fine-tune parameters, ensuring the simulation outcomes align more closely with observations (e.\,g. experimental measurements or high-fidelity simulations). This process involves addressing inverse problems to deduce model inputs, such as parameters, closure model constants, and initial or boundary conditions from the model output, treating the simulation as the forward problem. The goal is not to devise a universal closure model, but to calibrate a particular simulation configuration using (sparse) reference data. This technique is notably applied in fields like numerical weather forecasting (cf. \cite{kwon18}).

Data assimilation is split into two main strategies: variational and statistical. Variational DA aims at optimizing a cost function to bridge the gap between simulated outcomes and observational data by parameter adaptation. Conversely, statistical DA focuses on minimizing outcome variability. Nudging, a simpler form of DA, was originally devised for weather forecasts~\cite{asch16} and was more recently applied to turbulent flows~\cite{zauner_nudging-based_2022}. In scenarios with large sets of parameters, the computational demands of statistical methods, like the Kalman filter, become computationally expensive~\cite{singh16}. Here, the adjoint method presents an efficient variational DA solution as its computational expense is not tied to the number of parameters.

The adjoint method can be derived either in a continuous or a discrete way. In the continuous adjoint method, the primary step involves differentiating the governing equations, followed by discretization of the derived sensitivity equations for numerical solution. On the other hand, the discrete approach begins with the discretization of the governing equations, after which differentiation is performed to obtain sensitivity equations that are then solved numerically~\cite{he18a}.

Recent works have shown promising results in optimizing RANS simulations of stationary flows using sparse data through variational DA, enabling the reconstruction of mean flow profiles~\cite{foures14,brenner22,brenner23,li22,patel_turbulence_2023}.

DA for unsteady flows, however, remains a major challenge. Dynamic DA approaches like the Ensemble Kalman Filter (EnKF) or four-dimensional variational (4DVar) DA incur huge computational cost, since the flow dynamics cannot be neglected over a time interval during which observations are assimilated. EnKF have been extensively applied in the context of meteorology, and more recently also in the field of fluid mechanics~\cite{MELDI2017207}. In case of 4DVar DA, the adjoint is not simply the transpose of a matrix, but also the ``transpose'' of the model/operator dynamics~\cite{asch16}. In the past years, however, different attempts to reduce the computational cost of 4DVar DA for turbulent flows were made. Chandramouli~\etal.~\cite{chandramouli_4d_2020} reconstructed a time resolved wake flow in the transitional regime from measurements on two orthogonal 2D planes. To reduce the computational cost, the authors introduced a dynamics error model. He~\etal.~\cite{he_four-dimensional_2024} adopted a DA strategy using weak-constraint 4DVar DA to achieve super-temporal-resolution reconstruction of a turbulent jet beyond the Nyquist limit from low-sampling-rate observations. They utilized highly resolved LES data to generate synthetic measurements for validation, implementing a segregated assimilation procedure to separately assimilate initial condition, inflow boundary condition, and model error, demonstrating enhanced recovery of small-scale turbulence structures. Li~\etal.~\cite{li_unsteady_2023} introduced a continuous sliding window based weak-constraint 4DVar DA approach to enhance the unsteady flow over an airfoil using sparse spatiotemporal velocity observations. They simultaneously correct initial condition, boundary condition, and model form uncertainties through a spatiotemporally varying additive force. These works demonstrated that with some simplifications, 4DVar DA can in principle be adopted as a DA approach for turbulent flows, but the associated computational cost would be very high.

Therefore, Koltukluoğlu~\cite{shen_fourier_2019} tackled the problem by performing an adjoint-based DA in the Fourier domain using the harmonically balanced Navier--Stokes (HBNS) equations. The author demonstrated for blood flow simulations that the wall clock times could be substantially reduced. Rigas and Schmid~\cite{rigas22} also adopt the HBNS equations to formulate a data-driven framework to model the residual terms for the frequency-truncated HBNS equations. The authors show for two-dimensional cylinder flow at low Reynolds numbers that the computational cost could be reduced by solving for the coarse-grained dynamics. However, in both works, a harmonic solver is needed.

Symon~\etal.~\cite{symon_mean_2020} exploit three-dimensional variational (3DVar) DA to obtain the mean flow from sparse data, while the fluctuations of the unsteady flow are educed from resolvent analysis using time-resolved data at one point in the domain. Similarly, Franceschini~\etal.~\cite{franceschini_mean-_2021} demonstrated that the mean and unsteady flow can be reconstructed from sparse time-resolved data. They focus on transitional laminar flows, where the Fourier and resolvent modes can be aligned over energetic frequencies. This serves as a reduced order model for the unsteady flow and the amplitudes of the flow fluctuations can be tuned in the frequency space. Mons~\etal.~\cite{mons_optimal_2017} further elaborate on an optimal sensor placement procedure within a variational DA framework for unsteady flows. Furthermore, Plogmann~\etal.~\cite{plogmann23} time-averaged the URANS equations and performed 3DVar DA for time-averaged velocity data, which allowed for mean flow reconstruction and an improved vortex shedding frequency of different turbulent wake flows at a low computational cost. 

In the present study we introduce a novel approach exploiting the 3DVar DA approach to enhance the accuracy of URANS-based turbulent flow simulations. Our work is particularly focused on flows exhibiting time-periodic behavior, such as persistent vortex shedding, where the velocity signal is modeled using a Fourier series expansion. We adopt a stationary discrete adjoint method, wherein a time-discrete Fourier transformation of the URANS equations is performed. This allows for the optimization of stationary parameters in the Fourier-transformed URANS equations, expressed as a Fourier series expansion of the RANS model corrective force. The methodology is initially derived to accommodate $k$\footnote{Please note that at this point $k$ is not to be confused with the turbulent kinetic energy.} modes, yet we introduce a truncation after the first Fourier mode to showcase the fundamental capabilities of our framework.

We have opted for the discrete method due to its simplicity in constructing and implementing the discrete adjoint equations, as well as due to its natural handling of boundary conditions \cite{wang19}. Our optimization framework leverages an efficient semi-analytical approach for the computation of the cost function gradient within the discrete adjoint method, facilitating the evaluation of the derivative of the residual with respect to the parameters. We investigate the turbulent flow around a two-dimensional circular cylinder at a Reynolds number of 3900, using sparse synthetic reference data for both the zeroth and first Fourier modes. This choice is driven by our objective to demonstrate the framework's capabilities rather than achieving high-fidelity reconstruction, thus allowing for the use of a coarse computational mesh to reduce computational cost.

To the best of our knowledge, this is the first work to demonstrate that assimilating sparse Fourier transformed velocity reference data into the Fourier transformed URANS equations significantly improves the URANS simulation's predictive capabilities. Notably, the assimilation of zeroth Fourier mode data enables accurate mean flow reconstruction and enhances the prediction of vortex shedding frequency (Strouhal number), while the assimilation of first Fourier mode data augments the simulation's ability to capture low-frequency dynamics of the flow.

A key advantage of our approach is its circumvention of the computationally expensive 4DVar DA method, which necessitates a time-stepping scheme for assimilating time-resolved data. Instead, our more cost-effective 3DVar DA scheme allows for the assimilation of stationary (Fourier transformed) data sets, without the need for harmonic solvers. In particular, $2k+1$ stationary data sets represent $k$ modes of the flow. This not only reduces computational cost, but also decouples the time step in the unsteady forward problem from the time-resolved data resolution, allowing for the use of standard fluid solvers that operate in physical space and time.

The remainder of the paper is organized as follows. The method is introduced in section \ref{sec:methods} and results of unsteady flow around a circular cylinder are discussed in section \ref{sec:results}. Finally, in section \ref{sec:conclusion}, the work is summarized and future developments are suggested.


\section{Methods}
\label{sec:methods}

In this section it is elaborated on the unsteady simulation of incompressible flow problems. Additionally, a time-discrete Fourier transform is introduced to apply the stationary discrete adjoint method in the context of data assimilation.

\subsection{Unsteady Reynolds-averaged Navier--Stokes equations}
\label{sec:Unsteady Reynolds-averaged Navier--Stokes equations}

Turbulent flows, such as wake flows, often are unsteady, and often they are simulated using the URANS equations. Reynolds decomposing the quantity $\xi$ into its average $\bar{\xi}$ and fluctuation $\xi'$, applied to the velocity components $u_i$ and pressure $p$ in the momentum equation, yields the URANS euation
\begin{equation}
    \label{eq:urans_momentum_no_assumption}
    \frac{\partial \bar{u}_{i}}{\partial t}
    +
    \frac{\partial \bar{u}_{i} \bar{u}_{j}}{\partial x_{j}}
    +
    \frac{\partial}{\partial x_{i}}
    \left[
        \frac{\bar{p}}{\rho}
    \right]
    -
    \frac{\partial}{\partial x_{j}}
    \left[
        2 \nu \bar{S}_{ij}
    \right]
    +
    \frac{\partial \overline{u'_{i} u'_{j}} }{\partial x_{j}}
    = 0
\end{equation}
with constant density $\rho$ and the mean rate-of-strain tensor
\begin{equation}
    \bar{S}_{ij}
    =
    \frac{1}{2}
    \left(
        \frac{\partial \bar{u}_{i}}{\partial x_{j}}
        +
        \frac{\partial \bar{u}_{j}}{\partial x_{i}}
    \right)
    \, .
\end{equation}
Here the Reynolds stresses are modeled using the Boussinesq hypothesis, i.\,e.,
\begin{equation}
    \overline{u_i'u_j'} = \frac{2}{3}k \delta_{ij} - 2 \nu_t \bar{S}_{ij}
\end{equation}
with the eddy viscosity $\nu_t$.

\subsubsection{Data assimilation parameter}
\label{sec:Data assimilation parameter}

To account for discrepancies in the divergence of the modeled Reynolds stresses, we introduce a corrective force $f_{i}$ such that
\begin{equation}
    \frac{\partial \overline{u'_{i} u'_{j}}}{\partial x_{j}}
    =
    \frac{\partial}{\partial x_{j}}
    \left(
        \frac{2}{3}k\delta_{ij}
        -
        2\nu_{t}\bar{S}_{ij}
    \right)
    -
    f_{i}
    \, ,
\end{equation}
which is then subjected to a Stokes--Helmholtz decomposition, similarly done in~\cite{sliwinski_mean_2023,foures14,li22,patel_turbulence_2023,li_unsteady_2023,perot_turbulence_1999}, i.\,e.,
\begin{equation}
    f_{i}
    =
    -\frac{\partial \phi}{\partial x_{i}}
    +
    \epsilon_{ijk} \frac{\partial \psi_{k}}{\partial x_{j}}
    \, ,
\end{equation}
with the scalar potential $\phi$, the vector potential $\psi_k$, and the Levi-Civita symbol $\epsilon_{ijk}$.

The residual of the URANS momentum equation thus reads
\begin{equation}
    \label{eq:rans_momentum}
    \bar{R}
    =
    \frac{\partial \bar{u}_{i}}{\partial t}
    +
    \frac{\partial \bar{u}_{i} \bar{u}_{j}}{\partial x_{j}}
    +
    \frac{\partial p^{*}}{\partial x_{i}}
    -
    \frac{\partial}{\partial x_{j}}
    \left[
        2\nu_{\mathrm{eff}}\bar{S}_{ij}
    \right]
    -
    \epsilon_{ijk} \frac{\partial \psi_{k}}{\partial x_{j}}
    =
    0
\end{equation}
with the effective viscosity
\begin{equation}
    \nu_{\mathrm{eff}}
    =
    \nu
    +
    \nu_{t}
\end{equation}
and the modified pressure
\begin{equation}
    \label{eq:rans_pressure_mod}
    p^{*}
    =
    \frac{\bar{p}}{\rho}
    +
    \frac{2}{3} k
    +
    \phi
\end{equation}
that absorbs the averaged pressure $\bar{p}$, the isotropic part of the Reynolds stress tensor and the scalar potential $\phi$.

Please note that for the sake of simplicity $\bar{p}$ is used for the modified pressure $p^{*}$ in the remainder of the manuscript. Therefore only the time-dependent vector potential $\psi_k$ explicitly appears in the momentum equation. In our approach, data assimilation directly acts on $\psi_k$ and no additional equations need to be solved for $\phi$ nor $\psi_k$. For two-dimensional simulations complexity is further reduced, since only the $\psi_{z}$-component is non-zero. For more details it is referred to \cite{brenner23}.

\subsubsection{Fourier transform of the URANS equations}
\label{sec:Fourier transform of the URANS equations}

This work aims at extending the application of the discrete adjoint method (as discussed by Brenner~\etal.~\cite{brenner23}) for unsteady flow problems. Building upon the foundation laid by Plogmann~\etal.~\cite{plogmann23}, where time-averaging of the URANS equations was utilized to incorporate the stationary discrete adjoint method alongside time-averaged reference data, this study seeks further advancements. Plogmann~\etal.'s approach~\cite{plogmann23} enables the optimization of a stationary divergence-free force within the unsteady equations, facilitating the reconstruction of mean flow and improving the prediction the vortex shedding frequency (Strouhal number). Nonetheless, the approach is limited due to the utilization of a single stationary optimization parameter field, which restricts further improvement of the dynamic flow field. 

To account for this shortcoming, we introduce a Fourier transform of all quantities $\bar{\xi}$ in the URANS equations as
\begin{equation}
\label{eq:fourier_transform}
\hat{\xi}^k(\boldsymbol{x}) = \frac{1}{N} \sum_{j=0}^{N-1} \bar{\xi}(\boldsymbol{x},t_j) e^{-i \omega^{k} t_j}
\end{equation}
in a discrete manner (cf.~\cite{Pope2000}) with the corresponding inverse transform
\begin{equation}
\bar{\xi}(\boldsymbol{x},t)=\sum_{k=1-\frac{1}{2}N}^{\frac{1}{2}N} \hat{\xi}^{k}(\boldsymbol{x}) e^{i \omega^{k} t} \ ,
\end{equation}
where the angular frequency $\omega^k$ of mode $k$ is given by 
\begin{equation}
    \omega^{k}=\frac{2 k \pi}{T}=\frac{2 k \pi}{N \Delta t} 
    \, ,
\end{equation}
with the time period $T$ and $N$ equidistant samples at $t_j = j \Delta t$. Therefore it is assumed that the unsteady flow encounters a time-periodic behavior as it is the case for the flow around a cylinder with persistent vortex shedding. Moreover, it is assumed that the period $T$ is a known variable, since time-resolved data (from measurements or numerical experiments such as DNS or LES) needs to be Fourier transformed in order to serve as reference data.

Thus, the Fourier transformed URANS equations for mode $k$ read
\begin{equation}
    \label{eq:fourier_urans_momentum}
    \hat{R}^k_l
    =
    i \omega^k \hat{u}_{l}^k
    +
    \frac{\partial \widehat{\bar{u}_{l}\bar{u}_{j}}^k}{\partial x_{j}}
    +
    \frac{\partial \hat{p}^{k}}{\partial x_{l}}
    -
    \frac{\partial}{\partial x_{j}}
    \left[
        2\widehat{\nu_{\mathrm{eff}}\bar{S}_{lj}}^k
    \right]
    -
    \epsilon_{ljm} \frac{\partial \hat{\psi}_{m}^k}{\partial x_{j}}
    =
    0
    \, .
\end{equation}

For $k=0$ the Fourier transform is equivalent to a time-averaging operator. Hence, the work described in \cite{plogmann23} represents the simplest version of the proposed framework, i.\,e., a truncation after the zeroth mode. 

The Fourier transformed equations for the different modes are coupled due to the presence of nonlinear terms, in particular the convection and diffusion terms, hence a dependence between modes is evident.
 
\subsubsection{Truncated Fourier transformed URANS equations}
\label{sec:FT-URANS truncation}

For the sake of clarity we will only consider the zeroth (mean flow) and first modes (first harmonic). In section~\ref{sec:Harmonic components} we further elaborate on this choice for the flow under investigation. Recalling the assumption of a time-periodic flow solution, the real valued data assimilation parameter $\psi$ can be expressed by a Fourier series expansion as
\begin{equation}
    \psi = a + \sum_{k=1}^{N-1} \left[b^k \cos{\omega^k t} + c^k \sin{\omega^k t}\right] .
\end{equation}

Retaining only the zeroth and first modes of the Fourier series yields
\begin{equation}
\label{eq:data_assimilation_firstmode}
    \psi = a + b \cos{\omega^1 t} + c \sin{\omega^1 t}
\end{equation}
for the data assimilation parameter. Expressing the Fourier transformed URANS equations~\eqref{eq:fourier_urans_momentum} for the zeroth mode yields
\begin{equation}
    \label{eq:fourier_urans_momentum_0}
    \hat{R}^0_l
    =
    \frac{\partial \hat{u}^0_{l} \hat{u}_{j}^0}{\partial x_{j}}
    +
    \frac{\partial \hat{p}^{0}}{\partial x_{l}}
    -
    \frac{\partial}{\partial x_{j}} 
    \left[
        \left(
        \nu + \hat{\nu}^0_t
        \right)
        \left(
        \frac{\partial \hat{u}_l^0}{\partial x_{j}} + \frac{\partial \hat{u}_j^0}{\partial x_{l}}
        \right)
    \right]
    -
    \underbrace{\frac{\partial}{\partial x_j} \left( 2 \widehat{\nu_t'' \bar{S}_{jl}''}^0 \right)
    +
    \frac{\partial \widehat{\bar{u}_l''\bar{u}_j''}^0}{\partial x_j}}_{\mathrm{additional~stress~terms}}
    -
    \epsilon_{ljk} \frac{\partial a_k}{\partial x_{j}}
    =
    0
    \, ,
\end{equation}
while for the first mode they read
\begin{equation}
    \label{eq:fourier_urans_momentum_1}
    \begin{split}
        \hat{R}^1_l
        =
        i \omega^1 \hat{u}_{l}^1
        &+
        \frac{\partial}{\partial x_{j}} \left( \hat{u}_{l}^0 \hat{u}_{j}^1 + \hat{u}_{l}^1 \hat{u}_{j}^0 \right)
        +
        \frac{\partial \hat{p}^{1}}{\partial x_{l}}
        -
        \frac{\partial}{\partial x_{j}} 
        \left[
            \left(
            \nu + \hat{\nu}^0_t
            \right)
            \left(
            \frac{\partial \hat{u}_l^1}{\partial x_{j}} + \frac{\partial \hat{u}_j^1}{\partial x_{l}}
            \right)
            +
            \hat{\nu}^1_t
            \left(
            \frac{\partial \hat{u}_l^0}{\partial x_{j}} + \frac{\partial \hat{u}_j^0}{\partial x_{l}}
            \right)
        \right] \\
        &-
        \underbrace{\frac{\partial}{\partial x_j} \left( 2 \widehat{\nu_t'' \bar{S}_{jl}''}^1 \right)
        +
        \frac{\partial \widehat{\bar{u}_l''\bar{u}_j''}^1}{\partial x_j}}_{\mathrm{additional~stress~terms}} 
        -
        \epsilon_{ljk} \frac{\partial}{\partial x_{j}} \left( \frac{b_k}{2} - \frac{c_k}{2}i\right)
        =
        0
        \, .
    \end{split}
\end{equation}

Hence, the parameter $a$ only appears in eq.~\eqref{eq:fourier_urans_momentum_0} and the parameters $b$ and $c$ only in eq.~\eqref{eq:fourier_urans_momentum_1}. All other Fourier mode interactions caused by the Fourier transform of the nonlinear terms are summarized by $(\cdot)''$ in the additional stress terms. Discretizing these two residuals of eqs.~\eqref{eq:fourier_urans_momentum_0} and \eqref{eq:fourier_urans_momentum_1} in a coupled matrix system yields
\begin{equation}
\label{eq:coupled_fourier_momentum}
\hat{R} \left( \hat{U} \right)
=
\begin{bmatrix}
\hat{R}^0_\mathrm{R} \\
\hat{R}^0_\mathrm{I} \\
\hat{R}^1_\mathrm{R} \\
\hat{R}^1_\mathrm{I} \\
\end{bmatrix}
=
\begin{bmatrix}
\mathbf{A}^{00}_\mathrm{RR} & \mathbf{A}^{00}_\mathrm{RI} & \mathbf{A}^{01}_\mathrm{RR} & \mathbf{A}^{01}_\mathrm{RI} \\
\mathbf{A}^{00}_\mathrm{IR} & \mathbf{A}^{00}_\mathrm{II} & \mathbf{A}^{01}_\mathrm{IR} & \mathbf{A}^{01}_\mathrm{II} \\
\mathbf{A}^{10}_\mathrm{RR} & \mathbf{A}^{10}_\mathrm{RI} & \mathbf{A}^{11}_\mathrm{RR} & \mathbf{A}^{11}_\mathrm{RI} \\
\mathbf{A}^{10}_\mathrm{IR} & \mathbf{A}^{10}_\mathrm{II} & \mathbf{A}^{11}_\mathrm{IR} & \mathbf{A}^{11}_\mathrm{II} \\
\end{bmatrix}
\begin{bmatrix}
\hat{U}^0_\mathrm{R} \\
\hat{U}^0_\mathrm{I} \\
\hat{U}^1_\mathrm{R} \\
\hat{U}^1_\mathrm{I} \\
\end{bmatrix}
-
b_{\hat{U}}
=
\underbrace{\begin{bmatrix}
\mathbf{A}^0_{\hat{U}} & 0 & 0 & 0 \\
0 & 0 & 0 & 0 \\
\mathbf{A}^{10}_\mathrm{RR} & 0 & \mathbf{A}^0_{\hat{U}} & -\mathbf{A}^1_\omega \\
\mathbf{A}^{10}_\mathrm{IR} & 0 & \mathbf{A}^1_\omega & \mathbf{A}^0_{\hat{U}} \\
\end{bmatrix}}_{\mathbf{A}_{\hat{U}}}
\begin{bmatrix}
\hat{U}^0_\mathrm{R} \\
\hat{U}^0_\mathrm{I} \\
\hat{U}^1_\mathrm{R} \\
\hat{U}^1_\mathrm{I} \\
\end{bmatrix}
-
b_{\hat{U}}
=
0
\, ,
\end{equation}
where the complex equation for the first mode $\left( ^1 \right)$ is split into real $\left( _\mathrm{R} \right)$ and imaginary $\left( _\mathrm{I} \right)$ parts without loss of generality. The two additional terms are treated explicitly in the adjoint problem discretization. Consequently, the solution vector is defined as
\begin{equation}
\label{eq:forward_solution_fourier}
    \hat{U}
    =
    \left[
        \hat{U}^0_{\mathrm{R}}, \,
        \hat{U}^0_{\mathrm{I}}, \,
        \hat{U}^1_{\mathrm{R}}, \,
        \hat{U}^1_\mathrm{I}
    \right]^{T}
    \,  ,
\end{equation}
where, e.\,g. the real part of the zeroth mode is defined as
\begin{equation}
    \hat{U}^0_\mathrm{R}
    =
    \left[
        \hat{u}^0_\mathrm{R}, \,
        \hat{p}^0_\mathrm{R}
    \right]^{T}
    =
    \left[
        \hat{u}^0_{\mathrm{R},x}, \,
        \hat{u}^0_{\mathrm{R},y}, \,
        \hat{u}^0_{\mathrm{R},z}, \,
        \hat{p}^0_\mathrm{R}
    \right]^{T}
    \ , 
\end{equation}
which analogously holds for the remaining real and imaginary parts of the first mode. Therefore, each sub-system of eq.~\eqref{eq:coupled_fourier_momentum} for the real and imaginary components of the zeroth and first modes, respectively, also is discretized in a coupled manner. Thus, e.\,g. for the real component of the zeroth mode, the corresponding sub-system is discretized as
\begin{equation}
    \label{eq:coupled_residual}
    \hat{R}^0_\mathrm{R} 
    =
    \mathbf{A}^0_{\hat{U}}
    \hat{U}^0_\mathrm{R}
    -
    b_{\hat{U}^0_\mathrm{R}}
    =
    0
    \, ,
\end{equation}
which is analogously defined for the remaining Fourier modes. The coupling between the real and imaginary components of the first mode is denoted by $\mathbf{A}^1_{\omega}$. For the real component of the first mode, we write
\begin{equation}
    \label{eq:omega_contribution}
    \mathbf{A}^1_{\omega}
    \hat{U}_{\mathrm{R}}^1
    =
    \mathbf{A}^1_{\omega}
    \begin{bmatrix}
       \hat{u}_{\mathrm{R}}^1 \\
       \hat{p}_{\mathrm{R}}^1
    \end{bmatrix}
    =
    i \omega^1 \hat{u}_{\mathrm{R},i}^1
    \, ,
\end{equation}
since coupling only exists between the velocity components. Again, this discretization equivalently holds for the imaginary component of the first mode.

We recall that due to the non-linearity of the URANS equations there exist a coupling between the first and the zeroth modes, which is expressed in eq.~\eqref{eq:coupled_fourier_momentum} by the sub-matrices $\mathbf{A}^{10}_\mathrm{RR}$ and $\mathbf{A}^{10}_\mathrm{IR}$.

For a more in-depth discussion how the velocity and pressure coupling for each Fourier mode is treated in the coupled matrix system, please refer to~\cite{brenner23}.

\subsection{Data assimilation problem}
\label{sec:Data assimilation problem}

The data assimilation problem requires to minimize the discrepancy between the state variables computed by the URANS model and the existing reference data and thus can be constructed as an optimization problem. In this work the scalar cost function $f$ consists of a regularization function $f_{\hat{\psi}}$ and a discrepancy contribution $f_{\hat{U}}$, i.\,e.,
\begin{equation}
    \label{eq:cost_function}
    f\big(\hat{\psi}, \hat{U}\big)
    =
    f_{\hat{\psi}}\big(\hat{\psi}\big)
    +
    f_{\hat{U}}\big(\hat{U}\big)
    =
    \sum\limits_{k = 1 - \frac{1}{2}N}^{\frac{1}{2}N}
    f_{\hat{\psi}^k}\big(\hat{\psi}^k\big)
    +
    f_{\hat{U}^k}\big(\hat{U}^k\big)
    =
    \sum\limits_{k = 1 - \frac{1}{2}N}^{\frac{1}{2}N}
    f_k
    \approx
    f_0 + f_1
    \, ,
\end{equation}
when truncated after the first mode. It is subject to the residual $\hat{R}$ of the governing equation~\eqref{eq:coupled_fourier_momentum}, that is, one seeks
\begin{subequations}
    \begin{alignat}{2}
        \label{eq:minimization_problem}
        & \!\min_{\hat{\psi}}     & \quad & f\big(\hat{\psi}, \hat{U}\big) \\
        \label{eq:minimization_problem_constr_1}
        & \text{subject to} &       & \hat{R}\big(\hat{\psi}, \hat{U}\big) = 0 \ ,
    \end{alignat}
\end{subequations}
where $\hat{\psi}$ is the parameter to be optimized and $\hat{U}$ the Fourier transformed forward problem solution (cf.~eq.~\eqref{eq:forward_solution_fourier}).

The optimization involves an inverse problem, which is highly nonlinear, and usually underdetermined. Hence, a nonlinear optimization solver is used, but no assurance is given that there exists a unique solution. Therefore, some form of regularization has to be introduced to reduce the ambiguity (see~eq.~\eqref{eq:cost_function}), which is elaborated on in section~\ref{sec:Cost function and regularization}. For each data assimilation iteration, a parameter update is computed with a gradient descent approach. In the present work the optimization algorithm \emph{Demon Adam}~\cite{chen19} is chosen.

\subsubsection{Discrete adjoint method}
\label{sec:Discrete adjoint method}

To derive an expression for the cost function gradient, a Lagrangian
\begin{equation}
    \label{eq:lagrangian}
    \mathcal{L}\big(\hat{\psi}, \hat{U}\big)
    =
    f\big(\hat{\psi}, \hat{U}\big)
    -
    \lambda^{T} \hat{R}\big(\hat{\psi}, \hat{U}\big)
\end{equation}
with Lagrange multiplier
\begin{equation}
    \label{eq:lambda}
    \lambda
    =
    \left[
        \lambda^0_{\mathrm{R}}, \,
        \lambda^0_{\mathrm{I}}, \,
        \lambda^1_{\mathrm{R}}, \,
        \lambda^1_{\mathrm{I}}
    \right]^{T}
    \, 
\end{equation}
is introduced. The real part of the zeroth mode is
\begin{equation}
    \lambda_\mathrm{R}^0
    =
    \left[
        \lambda_{\hat{u}^0_\mathrm{R}}, \,
        \lambda_{\hat{p}^0_\mathrm{R}}
    \right]^{T}
    =
    \left[
        \lambda_{\hat{u}^0_{\mathrm{R},x}}, \,
        \lambda_{\hat{u}^0_{\mathrm{R},y}}, \,
        \lambda_{\hat{u}^0_{\mathrm{R},z}}, \,
        \lambda_{\hat{p}^0_\mathrm{R}}
    \right]^{T}
\end{equation}
and can be analogously written for all remaining Lagrange multipliers (i.\,e., for the real and imaginary component of the first mode).

The corresponding gradient with respect to the parameters $\hat{\psi}$ is derived as
\begin{equation}
    \label{eq:adjoint_gradient}
    \frac{\mathrm{d} f}{\mathrm{d} \hat{\psi}}
    =
    \frac{\partial f_{\hat{\psi}}}{\partial \hat{\psi}}
    -
    \lambda^{T} \frac{\partial \hat{R}}{\partial \hat{\psi}}
    \, .
\end{equation}
Rearranging the terms and considering that the derivative of the linearized forward problem residual $\hat{R}$ with respect to the Fourier transformed and linearized forward problem solution $\hat{U}$ corresponds to the respective system matrix $\mathbf{A}_{\hat{U}}$ (cf.~eq.~\eqref{eq:coupled_residual} and~\cite{brenner22,brenner23}) and yields
\begin{equation}
    \label{eq:coupled_adjoint}
    \left(\frac{\partial \hat{R}}{\partial \hat{U}}\right)^T \lambda 
    \approx
    \mathbf{A}_{\hat{U}}^{T} \, \lambda
    =
    \left( \frac{\partial f_{\hat{U}}}{\partial \hat{U}} \right)^{T}
    \, .
\end{equation}
While the right-hand side of this equation is analytically derived from the cost function and thus comes at low computational cost, solving this system of linear equations for $\lambda$ has a negligible computational cost compared to solving the forward problem, and is independent of the number of parameters $\hat{\psi}$.

Equation~\eqref{eq:coupled_adjoint} is also linearized in a coupled manner for the zeroth and first modes and reads
\begin{equation}
\label{eq:lagrangian_multiplier_coupled}
    \mathbf{A}_{\hat{U}}^T \lambda
    =
    \begingroup
    \renewcommand*{\arraystretch}{1.5}
    \begin{bmatrix}
        \begin{array}{cc|cc}
        \left(\mathbf{A}^0_{\hat{U}}\right)^T & 0 & \left(\mathbf{A}^{10}_\mathrm{RR}\right)^T & \left(\mathbf{A}^{10}_\mathrm{IR}\right)^T \\
        0 & 0 & 0 & 0 \\[0.5em]
        \hline
        0 & 0 & \left(\mathbf{A}^0_{\hat{U}}\right)^T & \left(\mathbf{A}^1_\omega\right)^T \\
        0 & 0 & -\left(\mathbf{A}^1_\omega\right)^T & \left(\mathbf{A}^0_{\hat{U}}\right)^T \\
        \end{array}
    \end{bmatrix}
    \begin{bmatrix}
        \begin{array}{c}
        \lambda^0_\mathrm{R} \\
        \lambda^0_\mathrm{I} \\[0.5em]
        \hline
        \lambda^1_\mathrm{R} \\
        \lambda^1_\mathrm{I} \\
        \end{array}
    \end{bmatrix}
    =
    \begin{bmatrix}
        \begin{array}{c}
        \left(\frac{\partial f_{\hat{U}}}{\partial \hat{U}^0_\mathrm{R}}\right)^T \\
        \left(\frac{\partial f_{\hat{U}}}{\partial \hat{U}^0_\mathrm{I}}\right)^T \\[0.5em]
        \hline 
        \left(\frac{\partial f_{\hat{U}}}{\partial \hat{U}^1_\mathrm{R}}\right)^T \\
        \left(\frac{\partial f_{\hat{U}}}{\partial \hat{U}^1_\mathrm{I}}\right)^T \\
        \end{array}
    \end{bmatrix}
    \endgroup
    \, , 
\end{equation}
where due to the transpose of $\mathbf{A}_{\hat{U}}$, a dependence (coupling) from the zeroth to the first mode now exists. To avoid solving such a big coupled matrix system, which can be numerically challenging and computationally expensive, we propose a sequential optimization. That is, first optimizing for the zeroth mode and in a second step, performing the optimization for the first mode. In section~\ref{sec:Cost function and regularization}, the definition of the cost function is introduced, which allows to decouple eq.~\eqref{eq:lagrangian_multiplier_coupled} and solve for $\lambda^0$ and $\lambda^1$ independently. A step-by-step explanation of the sequential optimization procedure is given in section~\ref{sec:Implementation}.

The derivatives $\partial \hat{R} / \partial \hat{\psi}$ of the Fourier transformed forward problem residual with respect to the parameters are evaluated using the approximate and efficient approach introduced by Brenner~\etal.~\cite{brenner22}.
In particular, the residual $\hat{R}$ is numerically linearized with respect to parameter $\hat{\psi}$ in \emph{OpenFOAM} as
\begin{equation}
    \hat{R}
    =
    \mathbf{A}_{\hat{\psi}} \, \hat{\psi}
    -
    b_{\hat{\psi}}
    =
    0
    \, .
\end{equation}

For the derivative with respect to parameter $\hat{\psi}$ that is needed for the evaluation of the adjoint gradient in eq.~\eqref{eq:adjoint_gradient}, this yields
\begin{equation}
    \frac{\partial \hat{R}}{\partial \hat{\psi}}
    =
    \frac{\partial}{\partial \hat{\psi}}
    \left[
        \mathbf{A}_{\hat{\psi}} \, \hat{\psi}
        -
        b_{\hat{\psi}}
    \right]
    =
    \mathbf{A}_{\hat{\psi}}
    \, .
\end{equation}

By expanding eq.~\eqref{eq:adjoint_gradient} with respect to the stationary parameters $a$, $b$ and $c$, the adjoint gradient, as derived in eq.~\eqref{eq:adjoint_gradient_derivation}, reads
\begin{equation}
\frac{\mathrm{d}f}{\mathrm{d}\hat{\psi}}
=
\begin{bmatrix}
\frac{\mathrm{d}f}{\mathrm{d}a} \quad \frac{\mathrm{d}f}{\mathrm{d}b} \quad \frac{\mathrm{d}f}{\mathrm{d}c}
\end{bmatrix}
=
    \begin{bmatrix}
    \left(\frac{\partial f}{\partial a} + \left(\lambda_{\mathrm{R}}^{0}\right)^T \mathbf{A}_{\hat{\psi}}\right) & \left(\frac{\partial f}{\partial b} + \frac{1}{2}\left(\lambda_{\mathrm{R}}^{1}\right)^T \mathbf{A}_{\hat{\psi}}\right)  &
    \left(\frac{\partial f}{\partial c} - \frac{1}{2}\left(\lambda_{\mathrm{I}}^{1}\right)^T \mathbf{A}_{\hat{\psi}}\right) 
    \end{bmatrix}
    \, .
\end{equation}

\subsubsection{Cost function and regularization}
\label{sec:Cost function and regularization}

As stated in eq.~\eqref{eq:cost_function}, the cost function consists of a regularization part and a discrepancy part. Here, total variation (TV) regularization is chosen to reduce ambiguity of the inverse problem. It is based on a measure for smoothness of the parameter field. Other works have focused on improving the regularization of inverse problems, since they always are underdetermined and thus ambiguous (e.\,g.~\cite{piroozmand_dimensionality_2023,epp22}). However, the optimal regularization method is not the focus of this work, so TV regularization as proposed in~\cite{brenner22} is considered sufficient. 

The discrepancy part of the cost function measures agreement of the Fourier transformed forward problem solution $\hat{U}$ with the reference data $\hat{U}^{\mathrm{ref}}$.
In the presented application only velocity data is assimilated, i.\,e.,

\begin{equation}
    \label{eq:discrepancy}
    \begin{split}
       f_{\hat{U}}\big(\hat{U}\big)
    &=
    \frac{
        1
    }{
        V^{\mathrm{ref}}
    }
    \sum\limits_{k = 1 - \frac{1}{2}N}^{\frac{1}{2}N}
    W^k
    \sum\limits_{j\in\mathcal{R}} \left[
        \sum\limits_{l\in\left\{x,y,z\right\}}
        \left(
            \hat{u}_{\mathrm{R},l,j}^k
            -
            \hat{u}_{\mathrm{R},l,j}^{k,\mathrm{ref}}
        \right)^{2}
        +
        \left(
            \hat{u}_{\mathrm{I},l,j}^k
            -
            \hat{u}_{\mathrm{I},l,j}^{k,\mathrm{ref}}
        \right)^{2}
    \right]
    V_{j} \\
    &=
    \sum\limits_{k = 1 - \frac{1}{2}N}^{\frac{1}{2}N}
    W^k f_{\hat{U}^k}
    \approx
    W^0 f_{\hat{U}^0} + W^1 f_{\hat{U}^1}
    \ , 
    \end{split} 
\end{equation}
where again the approximation involves a truncation after the first mode. Here, $k$ is the spectral mode, $W^k\in\left\{0,1\right\}$ a weight indicating which mode is considered, $\mathcal{R}$ is the list of reference cell indices $j$, $V_{j}$ the volume of cell $j$, and $V^{\mathrm{ref}}$ the volume of all reference cells. The cost function contribution of mode $k$ is therefore denoted by $f_{\hat{U}^k}$, where in this work only the zeroth and first modes are considered. Depending on the choice of $W_k$, only the zeroth or first mode may contribute to the cost function.

For two-dimensional flows, TV regularization only needs to be applied to the $\hat{\psi}_{z}$-component of the parameter field.
This is done without loss of generality, since the regularization can be applied to the other components as well (in 3D).
In particular, following the method by Brenner~\etal.~\cite{brenner22}, a function of the form
\begin{equation}
    \label{eq:regularization}
    f_{\hat{\psi}}\big(\hat{\psi}\big)
    =
    C^{\mathrm{reg}}
    \sum\limits_{l\in\Omega} \left[
        \frac{1}{\left|\mathcal{B}_{l}\right|}
        \sum\limits_{m\in\mathcal{B}_{l}} \left(
            \hat{\psi}_{z,l}
            -
            \hat{\psi}_{z,m}
        \right)^{2}
    \right]
\end{equation}
with weight parameter $C^{\mathrm{reg}}$ is used to punish non-smooth parameter fields.
Here, index $l$ loops over all cells in the simulation domain $\Omega$ and index $m$ loops over $\mathcal{B}_{l}$, the list of indices of neighboring cells of cell $l$, with $\left|\mathcal{B}_{l}\right|$ denoting the number of neighboring cells of cell $l$.

\subsubsection{Test function}
\label{sec:test function}

We introduce an additional functional $f_{\hat{U}}^\mathrm{test}$, denoted as the test function, which is formulated in a manner akin to the discrepancy cost function (cf.~eq.~\eqref{eq:discrepancy}). However, the key difference lies in the evaluation of $f_{\hat{U}}^\mathrm{test}$ over test data points, as opposed to reference data points. These test data points encompass all the points within the domain that are not categorized as reference points. This approach is frequently employed in the realm of machine learning, where a model's parameters are adjusted using training data. Following the phase of validation data analysis, test data serves to offer an impartial assessment of the model's performance after being trained. In the context of our discussion, the test function is instrumental in determining whether the optimization, as informed by reference data points, is effective across the broader domain.

\subsubsection{Sequential optimization procedure for zeroth and first Fourier modes}
\label{sec:Sequential optimization procedure for zeroth and first Fourier mode}

As pointed out in section~\ref{sec:Discrete adjoint method}, a sequential optimization by means of the different Fourier modes is proposed. Therefore, the optimization is performed for the zeroth and first Fourier modes independently. Below, the procedure is given with an explanation on how the adjoint equation~\eqref{eq:lagrangian_multiplier_coupled} is decoupled.

First, the modal weights in the cost function are set to $W^0 = 1$ and $W^1 = 0$ in eq.~\eqref{eq:cost_function}, such that only the zeroth Fourier mode is contributing to the discrepancy part of the cost function. Since $W^1 = 0$, the right hand side of eq.~\eqref{eq:lagrangian_multiplier_coupled} for the first Fourier mode is zero and the solution also yields zero for $\lambda_\mathrm{R}^1$ and $\lambda_\mathrm{I}^1$. Therefore, according to eq.~\eqref{eq:lagrangian_multiplier_coupled} the zeroth and first modes are no more coupled as the corresponding terms vanish.

In a second step the equation
\begin{equation}
\label{eq:lagrangian_multiplier_zero_mode}
    \left(\mathbf{A}^0_{\hat{U}}\right)^T 
    \lambda^0_\mathrm{R} 
    =
    \left(\frac{\partial f_{\hat{U}}}{\partial \hat{U}^0_\mathrm{R}}\right)^T  
\end{equation}
is solved for $\lambda_\mathrm{R}^0$ and $a$ is updated with $\mathrm{d}f / \mathrm{d}a$.

When the optimization in terms of the zeroth Fourier mode is completed, the data assimilation parameter $a$ will not be updated anymore. Now the parameters $b$ and $c$ are subjects of the optimization. Therefore, to optimize the first mode of the Fourier transformed velocity field, the following sequence is repeated until convergence:
First, the modal weights in the cost function are set to $W^0 = 0$ and $W^1 = 1$, such that only the first mode is contributing to the discrepancy part of the cost function.
Second, the system 
    \begin{equation}
\label{eq:lagrangian_multiplier_first_mode}
    \begin{bmatrix}
         \left(\mathbf{A}^0_{\hat{U}}\right)^T & \left(\mathbf{A}^1_\omega\right)^T \\
        -\left(\mathbf{A}^1_\omega\right)^T & \left(\mathbf{A}^0_{\hat{U}}\right)^T \\
    \end{bmatrix}
    \begin{bmatrix}
        \lambda^1_\mathrm{R} \\
        \lambda^1_\mathrm{I} \\
    \end{bmatrix}
    =
    \begin{bmatrix}
        \left(\frac{\partial f_{\hat{U}}}{\partial \hat{U}^1_\mathrm{R}}\right)^T \\
        \left(\frac{\partial f_{\hat{U}}}{\partial \hat{U}^1_\mathrm{I}}\right)^T \\
    \end{bmatrix}
\end{equation}
is solved for $\lambda_\mathrm{R}^1$ and $\lambda_\mathrm{I}^1$. Since $\left(\mathbf{A}_{\hat{U}}\right)^T$ in eq.~\eqref{eq:lagrangian_multiplier_coupled} is an upper triangular matrix, the solutions of $\lambda_\mathrm{R}^1$ and $\lambda_\mathrm{I}^1$ do not depend on $\lambda_\mathrm{R}^0$. Then $b$ and $c$ can be updated using $\mathrm{d}f / \mathrm{d}b$ and $\mathrm{d}f / \mathrm{d}c$, respectively. 

\subsection{Implementation}
\label{sec:Implementation}

The version \textit{foam-extend-5.0}~\cite{fe50} of the open-source field operation and manipulation platform, \textit{OpenFOAM}, known for its computational fluid dynamics (CFD) solvers, is utilized, leveraging the platform's pre-existing solvers and diverse capabilities. The approach adopted for solving both the forward and adjoint problems involves a fully coupled solution process. Notably, the \textit{transientFoam} solver was extended. The optimization itself is performed in \textit{Python} using the interface \textit{pyFOAM} to \textit{OpenFOAM}.

\begin{figure}[!ht]
    \centering
    \includegraphics[]{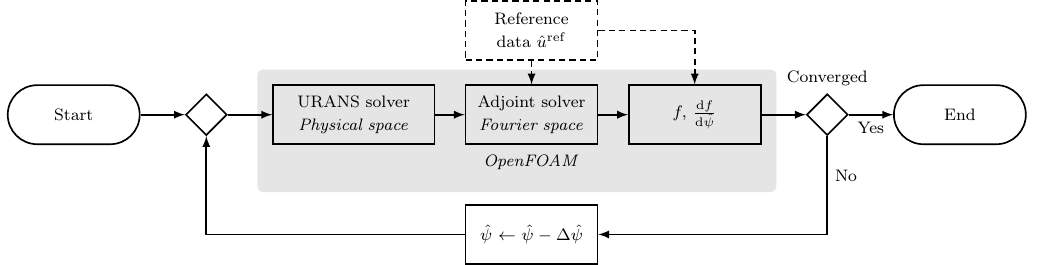}
    \caption{
        Flowchart of the optimization based data assimilation procedure.
        Starting from an initial URANS setup, the parameter field $\hat{\psi}$ is updated iteratively with $\Delta \hat{\psi}$ based on the cost function gradient. Depending on which Fourier mode is subject of the optimization, either parameter $a$ (zeroth mode) or $b$ and $c$ (first mode) are updated based on the respective cost function gradients. 
        A \textit{foam-extend-5.0} solver is used to evaluate the forward and adjoint problems, as well as the cost function and the gradients.
    }
    \label{fig:diagram_rans_da}
\end{figure}

Since a sequential optimization in terms of Fourier modes is targeted, the framework presented in~\cite{plogmann23} is adopted in this work. The iterative optimization procedure is depicted in fig.~\ref{fig:diagram_rans_da}. To construct the adjoint equations, Fourier transforming the URANS (forward problem) equation is required, which is performed during runtime for twenty periods of the periodic vortex shedding cycle. 

Details on the generation of the computational mesh, numerical schemes and solvers, optimization algorithm and further technical details are discussed in~\ref{app:Implementation}.


\section{Results and discussion}
\label{sec:results}

\subsection{Test case setup of flow around two-dimensional circular cylinder}
\label{sec:Test case setup of flow around two-dimensional circular cylinder}

\begin{figure}[!ht]
    \centering
    \includegraphics{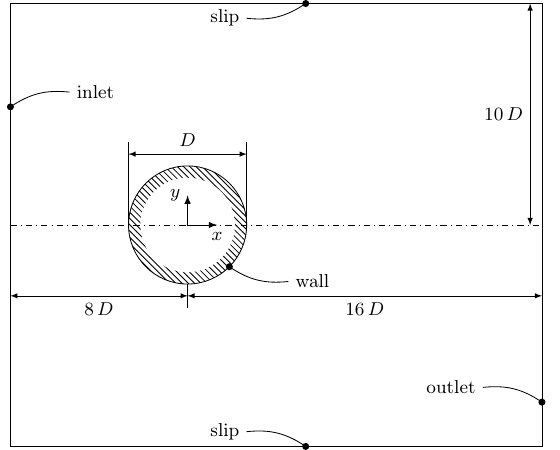}
    \caption{
            Simulation domain of the two-dimensional circular cylinder setup (not to scale). Mean flow is in $x$ direction.
            All length scales are normalized by cylinder diameter $D$. Boundary conditions (BC's) for velocity and pressure are taken from \cite{lehmkuhl_low-frequency_2013}. For the turbulence quantities, wall functions are applied at the cylinder wall and 2D BC's are used in spanwise directions. At the inlet Dirichlet BC's are used and the turbulence intensity is set to a small value reproducing the inflow conditions from \cite{lehmkuhl_low-frequency_2013}. Neumann BC's are set at the outlet for the turbulence properties.
    }
    \label{fig:circular_cylinder_2d}
\end{figure}

The performance of our data assimilation approach is demonstrated for flow around a two-dimensional circular cylinder (e.\,g.~\cite{lehmkuhl_low-frequency_2013}). A sketch of the geometry and boundary conditions is provided in fig.~\ref{fig:circular_cylinder_2d}. Depending on the Reynolds number, flow detachments occur and vortex shedding prevails in the wake of the cylinder. All length scales are expressed relative to the cylinder diameter $D$ and the Reynolds number is computed from $D$, the free-stream velocity $u_\infty$, and the kinematic viscosity $\nu$. In this work we analyze the case for
\begin{equation*}
    \mathrm{Re}
    =
    \frac{u_{\infty}D}{\nu}
    =
    \num{3900}
\end{equation*}
with synthetic reference data generated by a stationary source term in the momentum equation. A time step size of $\Delta t = 0.1 D / u_\infty$ is used and the mesh consists of 5700 hexahedral elements with a ratio of the smallest to largest cell of $7.614 \cdot 10^{-3}$.

\subsection{Synthetic reference data}
\label{sec:Synthetic reference data}

The reference data is synthetically generated by adding the force  
\begin{equation}
    \label{eq:synthetic reference force}
     f_x (x,y) = 
    \begin{cases}
        0.01\sqrt{(x-1)^2+4y^2} -0.05, & \sqrt{(x-1)^2+4y^2} < 5 \\
        0, & \text{otherwise}
    \end{cases}
\end{equation}
to the $x$-component of the URANS momentum equations~\eqref{eq:urans_momentum_no_assumption} and
\begin{equation}
      f_y (x,y) = 
    \begin{cases}
        0.0175\sqrt{x^2+16(y-1)^2} -0.07, & \sqrt{x^2+16(y-1)^2} < 4 \\
        -0.0175\sqrt{x^2+16(y+1)^2} +0.07, & \sqrt{x^2+16(y+1)^2} < 4 \\
        0, & \text{otherwise}
    \end{cases}  
\end{equation}
to the $y$-component.

Based on the URANS simulation with the synthetic force, time-resolved velocity data is generated and Fourier transformed subsequently. We obtain three data sets for the zeroth and first modes. Namely, the real component of the zeroth mode $\hat{u}_{\mathrm{R}}^{0,\mathrm{ref}}$, the real component of the first mode $\hat{u}_{\mathrm{R}}^{1,\mathrm{ref}}$ and the imaginary component of the first mode $\hat{u}_{\mathrm{I}}^{1,\mathrm{ref}}$.

It would certainly be interesting to use reference data from the literature, e.\,g. DNS, high-resolution LES or from an experiment. However, the focus of this work is not on the optimization of the simulated flow based on real data, but on the demonstration of the novel framework for unsteady flows. In addition, in the existing literature reference data for the flow investigated here is only provided in time-averaged, but not in time-resolved form. This is of course related to the amount of data. In previous works~\cite{brenner22, brenner23, plogmann23}, however, the authors demonstrated that the discrete adjoint based data assimilation framework, adapted in this work, is robust and efficient for a series of test cases with different velocity reference data (synthetic, DNS and LES data) sources.

\subsection{Mean flow and harmonic components}
\label{sec:Harmonic components}

\begin{figure}[!ht]
     \centering\begin{subfigure}[t]{0.245\textwidth}
         \centering
         \includegraphics[width=\textwidth]{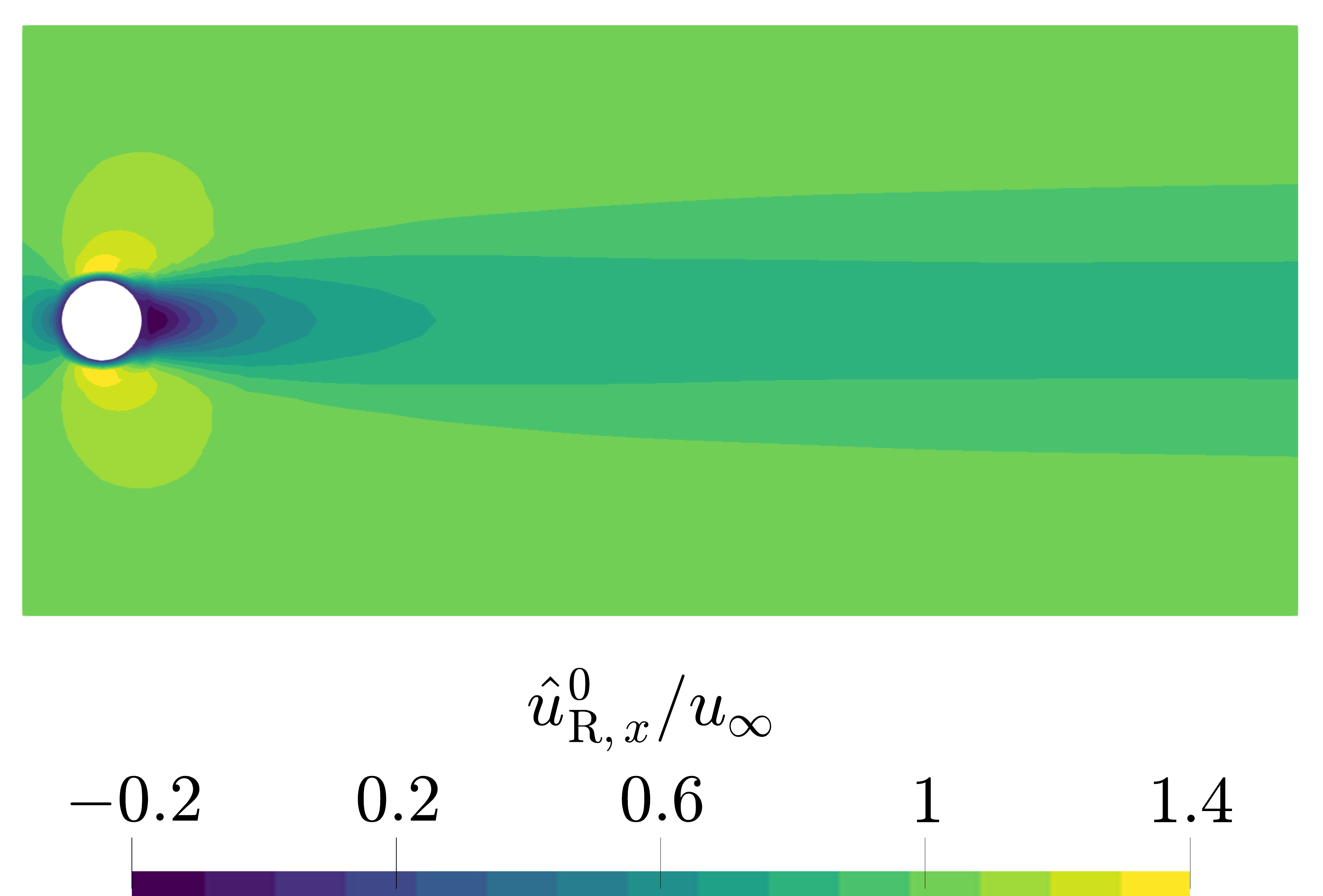}
     \end{subfigure}
     \hfill
     \begin{subfigure}[t]{0.245\textwidth}
         \centering
         \includegraphics[width=\textwidth]{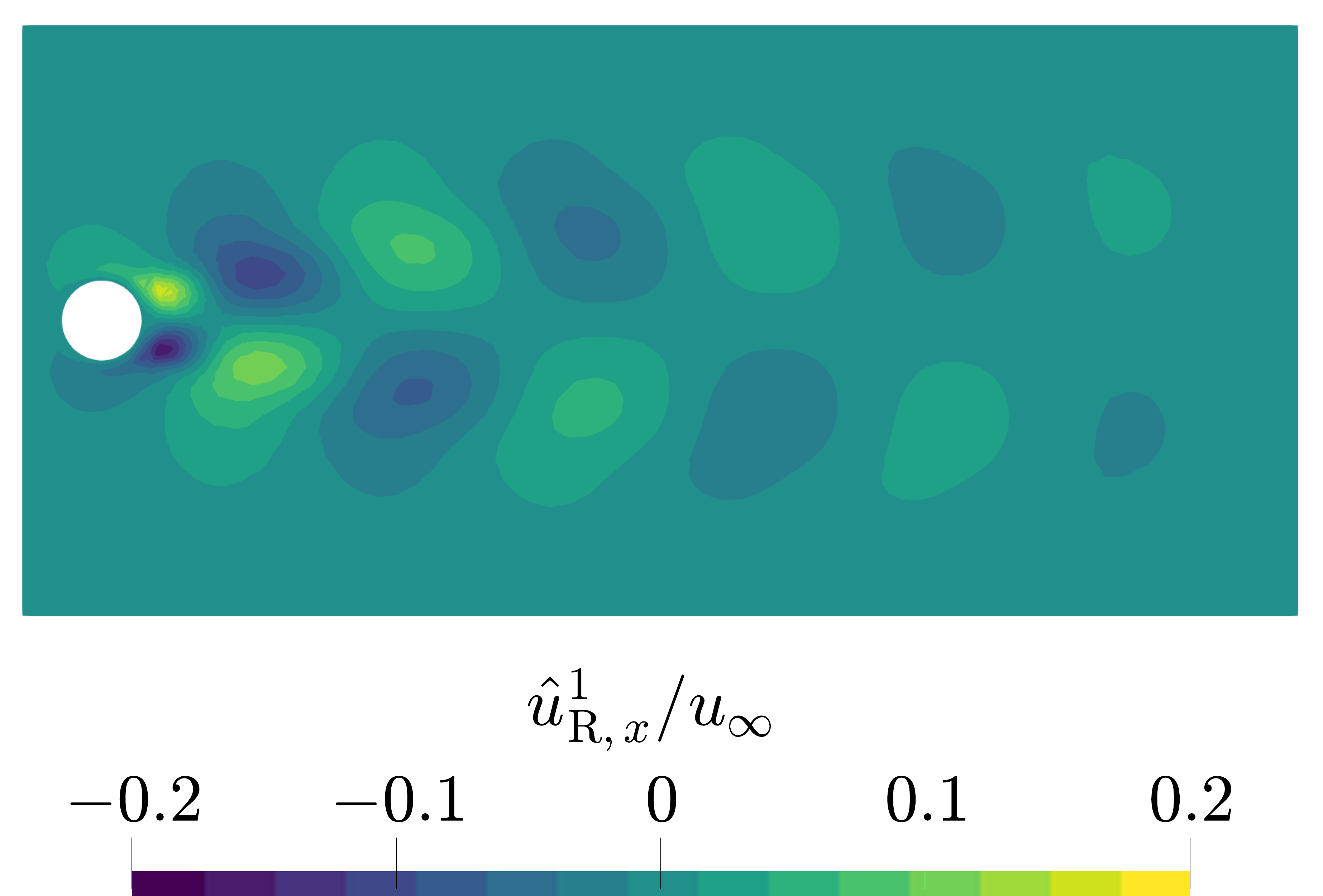}
     \end{subfigure}
     \hfill
     \begin{subfigure}[t]{0.245\textwidth}
         \centering
         \includegraphics[width=\textwidth]{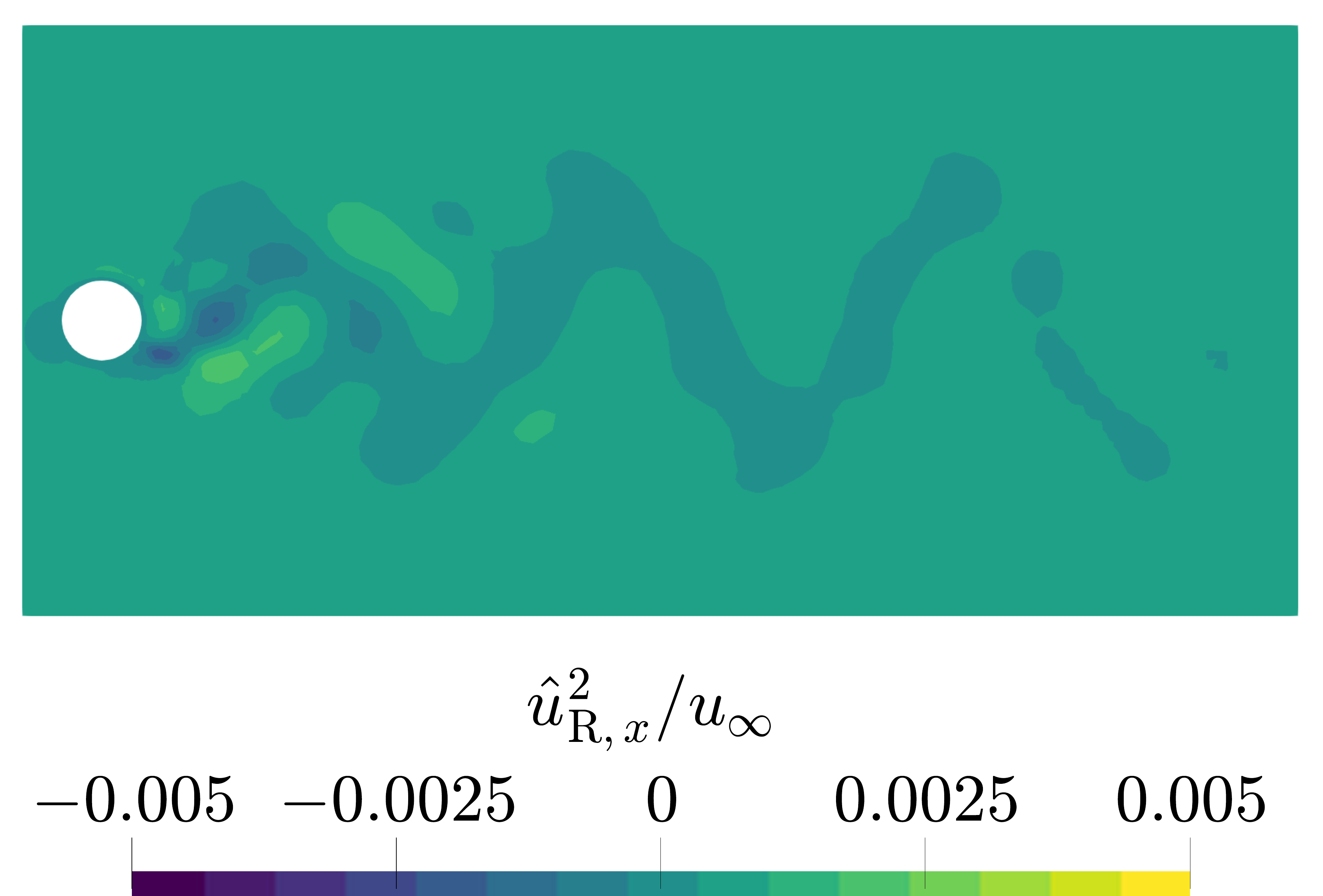}
     \end{subfigure}
     \hfill
     \begin{subfigure}[t]{0.245\textwidth}
         \centering
         \includegraphics[width=\textwidth]{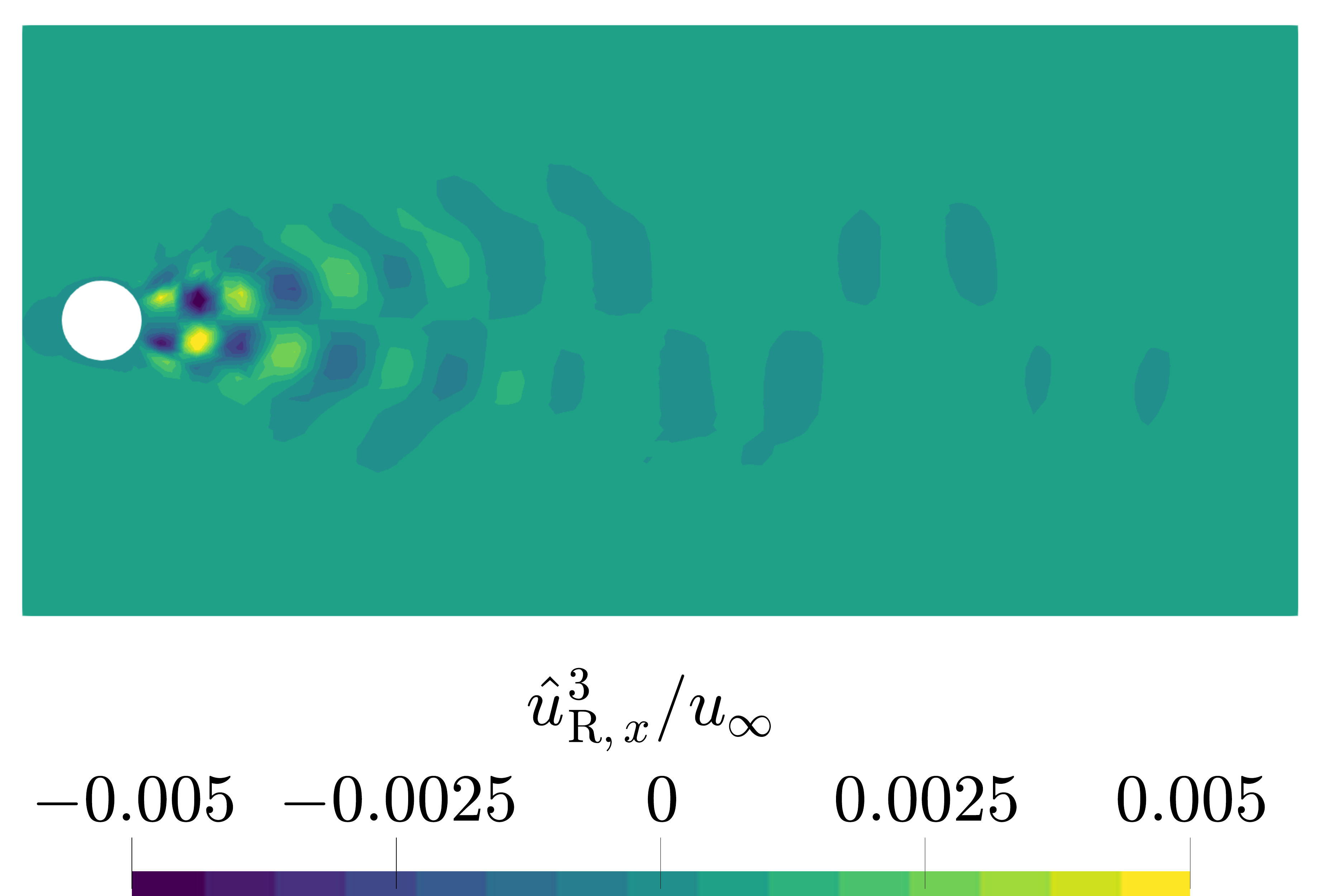}
     \end{subfigure}
     \hfill
     \begin{subfigure}[t]{0.245\textwidth}
         \centering
         \includegraphics[width=\textwidth]{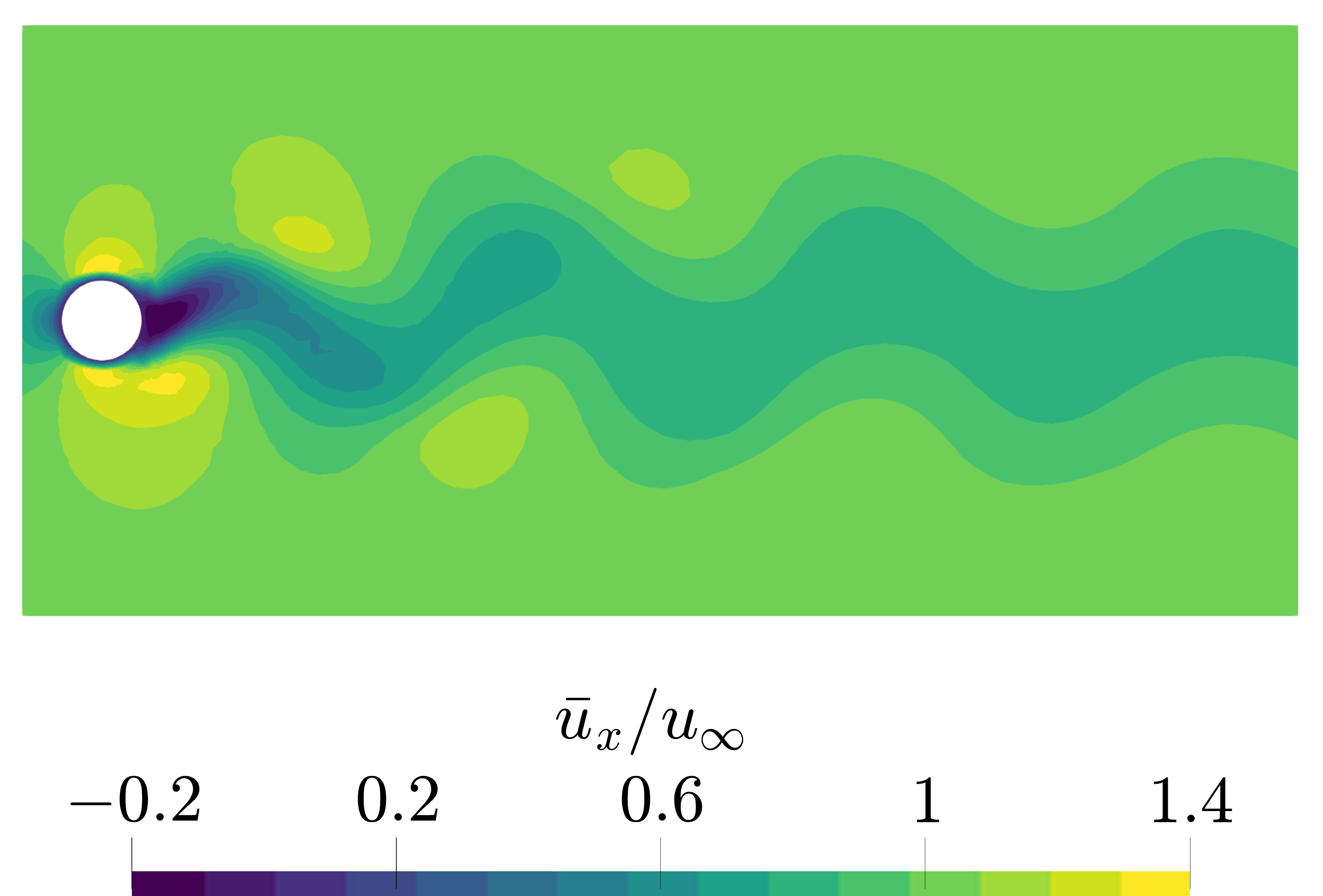}
     \end{subfigure}
     \hfill
     \begin{subfigure}[t]{0.245\textwidth}
         \centering
         \includegraphics[width=\textwidth]{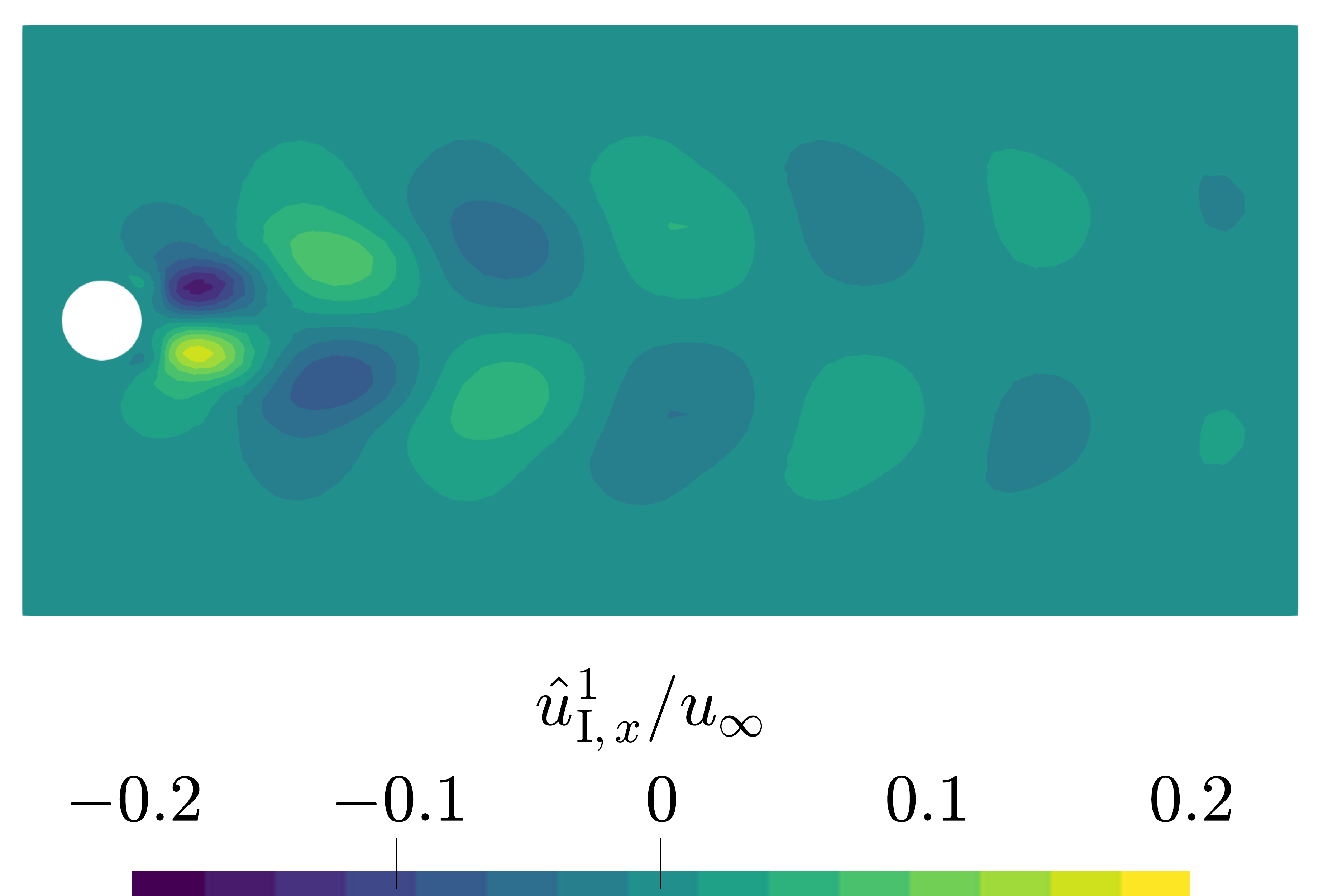}
     \end{subfigure}
     \hfill
     \begin{subfigure}[t]{0.245\textwidth}
         \centering
         \includegraphics[width=\textwidth]{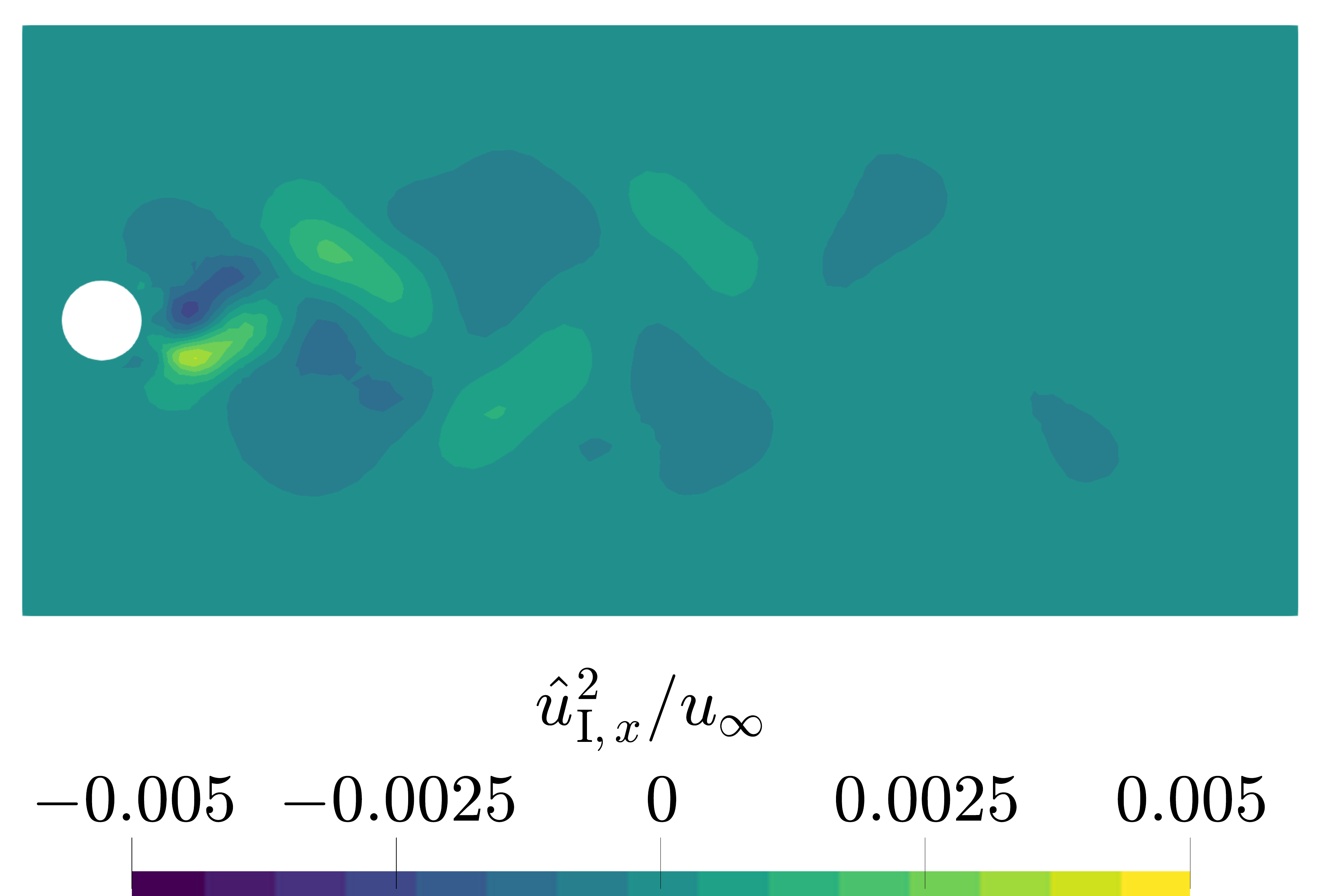}
     \end{subfigure}
     \hfill
     \begin{subfigure}[t]{0.245\textwidth}
         \centering
         \includegraphics[width=\textwidth]{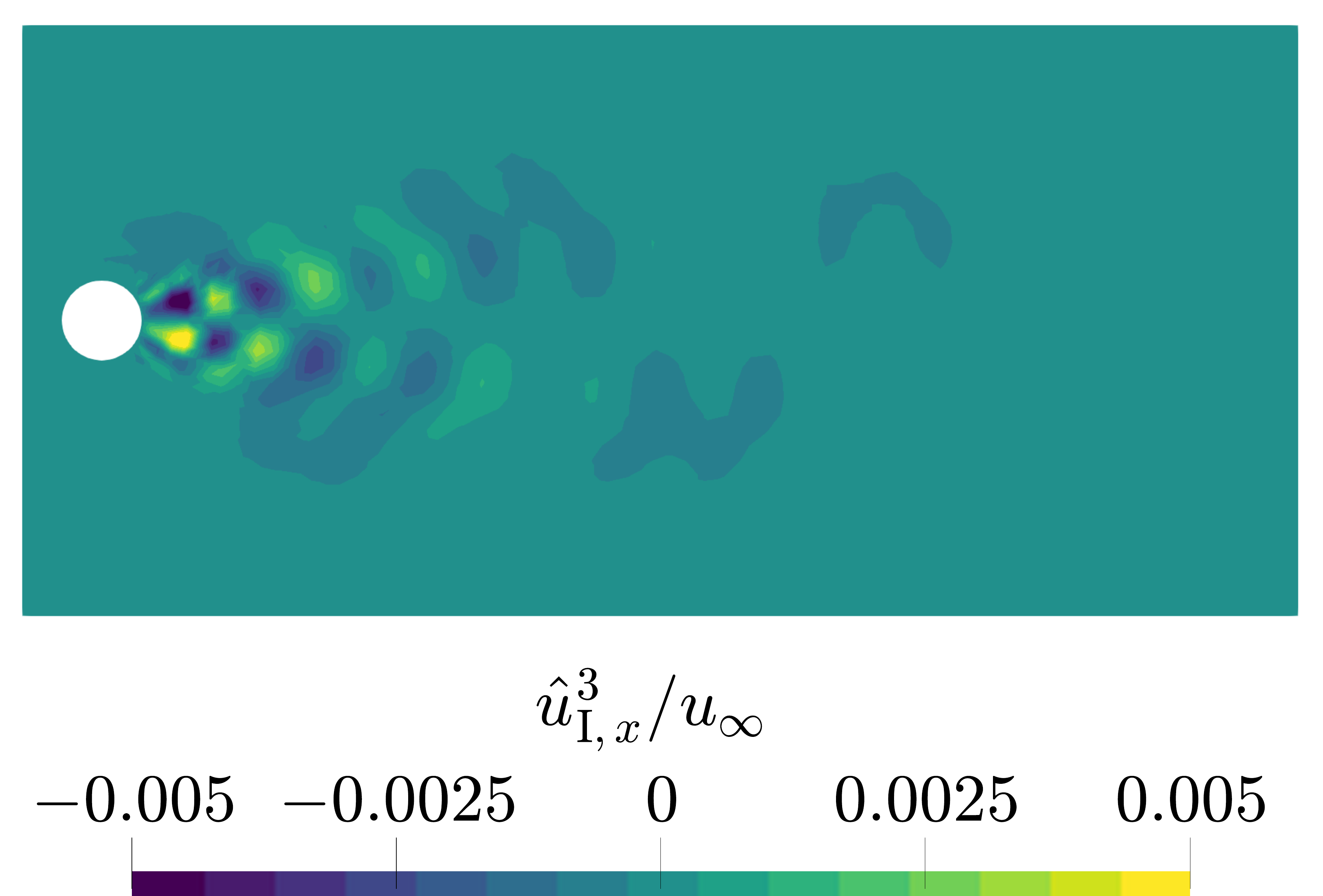}
     \end{subfigure}
\caption{Streamwise velocity components are shown for the flow around the two-dimensional circular cylinder at $Re = 3900$ obtained from the baseline URANS simulation. Mean flow and real harmonic components are presented in the top and imaginary harmonic components in the bottom line. The instantaneous velocity is depicted in the bottom left corner. The Strouhal number is $St = f D / u_\infty = 0.1975$. The Fourier transform was performed for 40 periods of $T u_\infty / D = u_\infty / (fD) = 1 / St = 5.063$. For better illustration, the velocity fields are shown for $-D \leq x \leq 15D$ and $-3.7D \leq y \leq 3.7 D$, respectively.}
\label{fig:harmonics}
\end{figure}

Figure~\ref{fig:harmonics} depicts the URANS baseline simulation results in terms of the instantaneous velocity field alongside the mean flow and the first three complex harmonic components for the streamwise velocity component. It is observed that higher harmonics are typically associated with higher frequencies, shorter wavelengths, and lower amplitudes, underscoring the flow dynamics they represent. Importantly, we would like to stress again that the URANS approach is resolving the deterministic and coherent vortex structures that manifest at a consistent vortex shedding frequency, which is crucial for capturing the essence of the wake flow's dynamics. However, it is worth noting that smaller vortices characterized by higher frequencies, represented by these higher harmonics, are beyond the resolution of URANS and thus are modeled. Consequently, the analysis predominantly focuses on assimilating the zeroth mode (mean flow) and the first harmonic/mode. These components are pivotal as they represent the most critical features of the flow, including the fundamental vortex shedding frequency, offering a comprehensive understanding of the present wake flow's characteristics.

\subsection{Mean flow reconstruction (zeroth Fourier mode)}
\label{sec:Mean flow reconstruction}

\begin{figure}[!ht]
     \centering
     \begin{subfigure}[t]{0.49\textwidth}
         \centering
         \includegraphics[width=\textwidth]{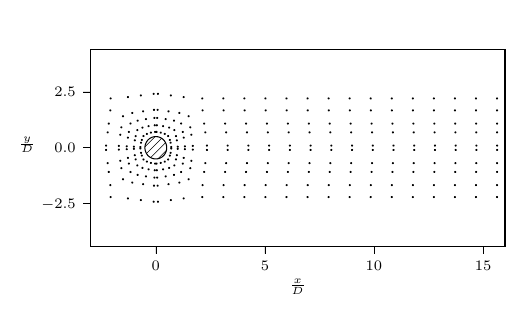}
         \caption{Reference data point locations around circular cylinder and its wake. In total 284 points are selected.}
         \label{fig:reference data points circular cylinder}
     \end{subfigure}
     \hfill
     \begin{subfigure}[t]{0.49\textwidth}
         \centering
         \includegraphics[width=\textwidth]{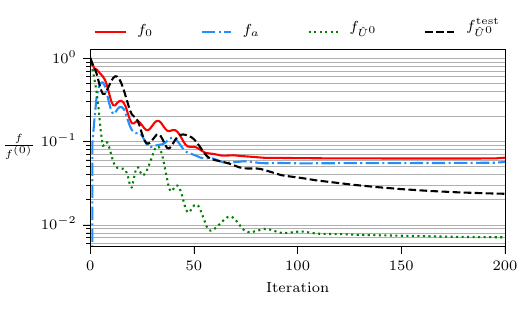}
         \caption{Normalized cost function of zeroth mode.}
         \label{fig:cost function circular cylinder mode 0 opt}
     \end{subfigure}
     \hfill
     \begin{subfigure}[t]{0.49\textwidth}
         \centering
         \includegraphics[width=\textwidth]{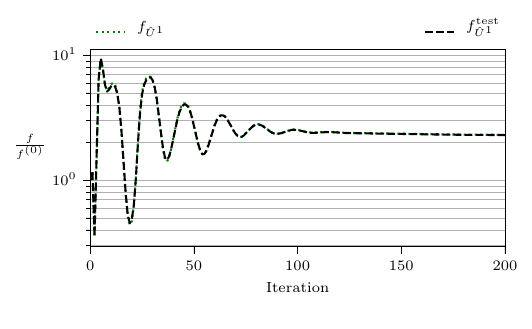}
         \caption{Normalized cost function of first mode.}
         \label{fig:cost function circular cylinder mode 1}
     \end{subfigure}
     \hfill
     \begin{subfigure}[t]{0.49\textwidth}
         \centering
         \includegraphics[width=\textwidth]{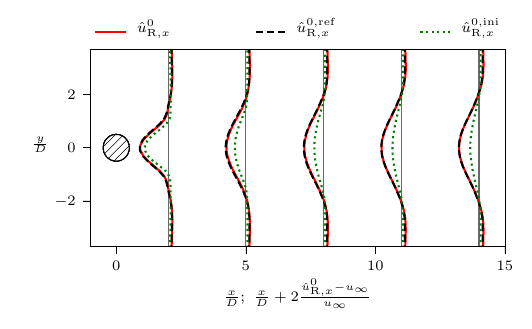}
         \caption{Mean (real component of zeroth mode) streamwise velocity profiles in the wake.}
         \label{fig:velocity profiles circular cylinder wake mode 0}
     \end{subfigure}
     \hfill
     \begin{subfigure}[t]{0.49\textwidth}
         \centering
         \includegraphics[width=\textwidth]{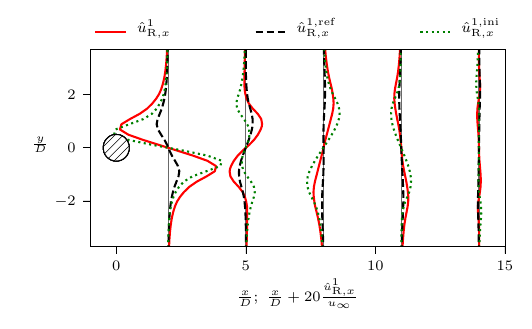}
         \caption{Profiles of the streamwise real component of the first Fourier mode of velocity in the wake.}
         \label{fig:velocity profiles circular cylinder wake mode 1, re after 0}
     \end{subfigure}
     \hfill
     \begin{subfigure}[t]{0.49\textwidth}
         \centering
         \includegraphics[width=\textwidth]{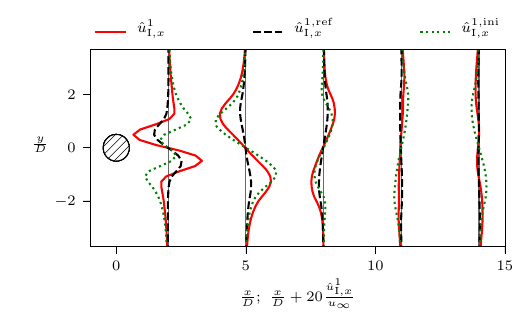}
         \caption{Profiles of the streamwise imaginary component of the first Fourier mode of velocity in the wake.}
         \label{fig:velocity profiles circular cylinder wake mode 1, im after 0}
     \end{subfigure}
        \caption{Optimization of mean flow (zeroth mode) around the two-dimensional circular cylinder using synthetic reference data. Optimization step size is $\eta=\num{5e-5}$ with a maximum number of optimization steps $N_\mathrm{opt}=200$. The forward problem solver is running for 20 periods of $T u_\infty / D = 1 / St = 5.57$. Optimization only is performed for the zeroth mode ($W^0=1$, $W^1=0$). Hence, the parameter field $a$ is updated but $b$ and $c$ are not active yet. The regularization weight parameter is set to $C^{\mathrm{reg}}=\num{2e-1}$. Compared are the streamwise velocity profiles from the baseline ($\hat{u}_{x}^\mathrm{ini}$), reference ($\hat{u}_{x}^\mathrm{ref}$) and optimized ($\hat{u}_{x}$) simulation for the real and imaginary component of the zeroth and first mode.}
        \label{fig:circular cylinder zeroth mode}
\end{figure}

As depicted in fig.~\ref{fig:reference data points circular cylinder}, 284 reference data points around the cylinder and in the wake are selected to demonstrate the efficacy of our proposed method. In a previous study~\cite{plogmann23}, Plogmann~\etal.~investigated the importance of near-cylinder reference data points for the improvement of the Strouhal number and also showed that reference data points in the wake are crucial for the mean flow reconstruction in that region. In terms of sparsity of the reference data, the generic optimization framework presented in~\cite{brenner22,brenner23} for RANS simulations, that we built upon, proved to be robust. Therefore, the scope of this work is not to study effects of the reference data placement, but to discuss the basic capabilities of the proposed data assimilation framework for unsteady flows.

In a first step, only an optimization in terms of the mean flow is performed. Thus, the weights in the cost function are set to $W^0=1$ and $W^1=0$, respectively. Optimizing only for the stationary parameter $a$ leads to a decrease in the discrepancy part of the cost function $f_{\hat{U}^0}$ of more than two orders of magnitude, as can be seen in fig.~\ref{fig:cost function circular cylinder mode 0 opt}. To this end, we would like to stress again that this cost function only expresses discrepancies of the velocity in terms of the zeroth mode. Due to the presence of regularization, all remaining points in the domain also show an improvement, that is, $f_{\hat{U}^0}^\mathrm{test}$ decreases substantially with the number of optimization steps. Meanwhile, the test function and discrepancy part of the cost function expressed in terms of the first mode of the Fourier transformed velocity slightly increases during the zeroth mode optimization (see~fig.~\ref{fig:cost function circular cylinder mode 1}). The optimized mean flow profiles in fig.~\ref{fig:velocity profiles circular cylinder wake mode 0} in the wake of the cylinder are in very good agreement with the reference profiles; unlike the baseline profiles. The streamwise velocity profiles also match well in the near-cylinder region due to the existence of near-cylinder reference data points, as reported in~\cite{plogmann23}. As extensively discussed in~\cite{plogmann23}, the Strouhal number improves by assimilating mean velocity data only. However, the existence of near-cylinder reference data points is crucial. Moreover, the degree of improvement in the Strouhal number strongly depends on how much the cost function decreases. Hence, the better the optimization in terms of the zeroth mode of velocity (mean flow) works, the better the agreement of reference and optimized Strouhal number is. In~\cite{plogmann23}, this was demonstrated for flows around three different geometries at three different Reynolds numbers in the turbulent regime. Here, the Strouhal number improves from $St^\mathrm{ini} = 0.1975$ to $St=0.1795$, which exactly corresponds to the reference ($St^\mathrm{ref}=0.1795$).

In figs.~\ref{fig:velocity profiles circular cylinder wake mode 1, re after 0} and \ref{fig:velocity profiles circular cylinder wake mode 1, im after 0}, one can see that the wake profiles of the first Fourier mode of the streamwise velocity did not improve but changed the shapes in comparison to the baseline result.

\subsection{Unsteady flow reconstruction (first Fourier mode)}
\label{sec:Unsteady flow reconstruction}

As explained in section~\ref{sec:Sequential optimization procedure for zeroth and first Fourier mode}, we aim for a sequential optimization in terms of the Fourier modes of the velocity field. Therefore we now seek for optimal parameters $b$ and $c$, while the parameter $a$ obtained from the zeroth mode optimization (see section~\ref{sec:Mean flow reconstruction}) is fixed. Hence, the weights in the cost function are set to $W^0=0$ and $W^1=1$.

\begin{figure}[!ht]
     \centering
     \begin{subfigure}[t]{0.49\textwidth}
         \centering
         \includegraphics[width=\textwidth]{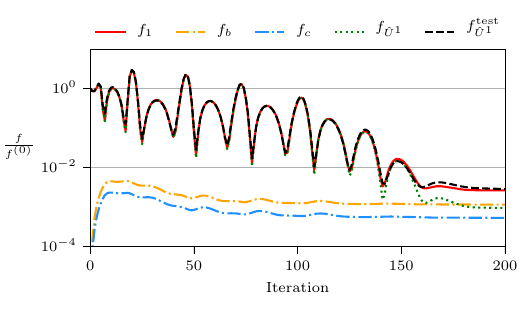}
         \caption{Normalized cost function of first mode during optimization.}
         \label{fig:cost function circular cylinder mode 1 opt}
     \end{subfigure}
     \hfill
     \begin{subfigure}[t]{0.49\textwidth}
         \centering
         \includegraphics[width=\textwidth]{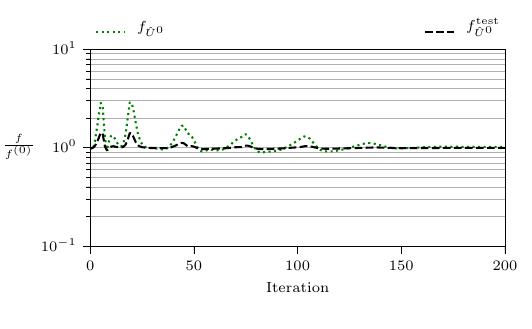}
         \caption{Normalized cost function of zeroth mode.}
         \label{fig:cost function circular cylinder mode 0}
     \end{subfigure}
     \hfill
     \begin{subfigure}[t]{0.49\textwidth}
         \centering
         \includegraphics[width=\textwidth]{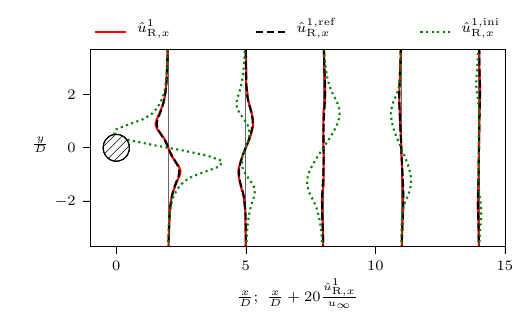}
         \caption{Profiles of the streamwise real component of the first Fourier mode of velocity in the wake.}
         \label{fig:velocity profiles circular cylinder wake mode 1 Re}
     \end{subfigure}
     \hfill
     \begin{subfigure}[t]{0.49\textwidth}
         \centering
         \includegraphics[width=\textwidth]{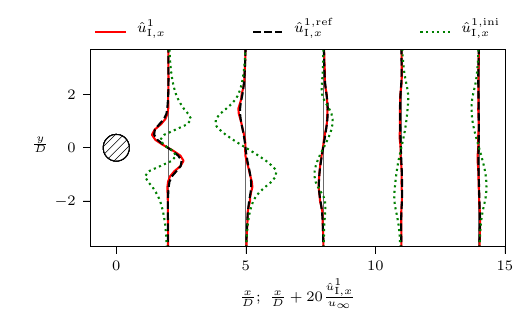}
         \caption{Profiles of the streamwise imaginary component of the first Fourier mode of velocity in the wake.}
         \label{fig:velocity profiles circular cylinder wake mode 1 Im}
     \end{subfigure}
        \caption{Optimization of flow dynamics (first mode) around the two-dimensional circular cylinder using synthetic reference data. Optimization step size is $\eta=\num{8e-6}$ for both parameters $b$ and $c$ with a maximum number of optimization steps $N_\mathrm{opt}=200$. The forward problem solver is running for 20 periods of $T u_\infty / D = 1 / St = 5.57$. Optimization only is performed for the first mode ($W^0=0$, $W^1=1$). Hence, the parameter fields $b$ and $c$ are updated, but $a$ is fixed at optimized result from zeroth mode optimization. The regularization weight parameter is set to $C^{\mathrm{reg}}=\num{8e-4}$ for both parameters $b$ and $c$.}
        \label{fig:circular cylinder results mode 1}
\end{figure}

As can be seen in fig.~\ref{fig:cost function circular cylinder mode 1 opt}, the optimization for the stationary parameters $b$ and $c$ leads to a decrease in the discrepancy part $f_{\hat{U}^1}$ of the cost function by three orders of magnitude. Similarly, the test function $f_{\hat{U}^1}^\mathrm{test}$ shows a significant improvement, which means that the optimization works well in the whole domain. It should be mentioned that the cost function values in figs.~\ref{fig:cost function circular cylinder mode 1 opt} and \ref{fig:cost function circular cylinder mode 0} are normalized with their first cost function value $f^{(0)}$ of the first Fourier mode optimization. These values correspond to the cost function values of $f^{(200)}$ from the zeroth Fourier mode optimization depicted in figs.~\ref{fig:cost function circular cylinder mode 0 opt} and \ref{fig:cost function circular cylinder mode 1}. Therefore the optimization of the first velocity Fourier mode does not deteriorate the optimized results of the zeroth Fourier mode, since the discrepancy cost function value for the zeroth mode during the optimization of the first mode is still the same after 200 optimization steps (see~fig.~\ref{fig:cost function circular cylinder mode 0}). Furthermore, the Strouhal number as well as the mean velocity profiles do not change anymore during the first Fourier mode optimization, as the discrepancy and test function stay unaltered. 

The first Fourier mode transformed velocity profiles highlight the drastic improvement of the optimization. Before optimization, a huge discrepancy is obvious for both the real (see~fig.~\ref{fig:velocity profiles circular cylinder wake mode 1, re after 0}) and the imaginary (see~fig.~\ref{fig:velocity profiles circular cylinder wake mode 1, im after 0}) components. However, after optimization the velocity profiles match very well (see figs.~\ref{fig:velocity profiles circular cylinder wake mode 1 Re} and \ref{fig:velocity profiles circular cylinder wake mode 1 Im}). In conclusion, the low-frequency dynamics of the flow associated with the frequency $\omega^1$  agree on average (Fourier transformed over twenty periods) very well with the reference. At the same time, the mean flow reconstruction together with the vortex shedding frequency (Strouhal number) are not degraded. Therefore, the sequential optimization for the zeroth and subsequently for the first mode proved to be very effective.

\subsection{Optimal data assimilation parameter fields}
\label{sec:Optimal data assimilation parameter fields}

\begin{figure}[!ht]
     \centering
     \begin{subfigure}[t]{0.32\textwidth}
         \centering 
         \includegraphics[width=\textwidth]{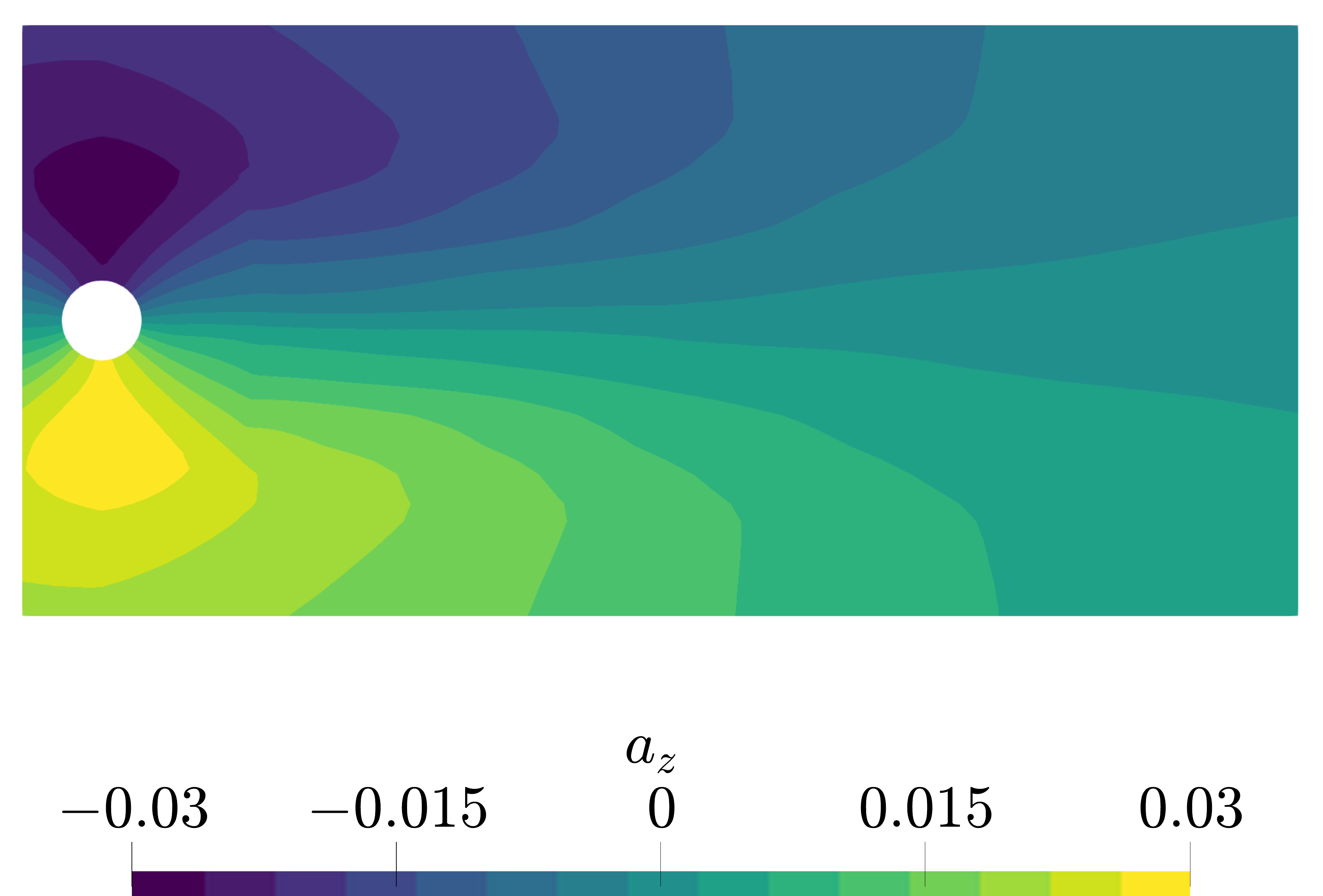}
     \end{subfigure}
     \hfill
     \begin{subfigure}[t]{0.32\textwidth}
         \centering
         \includegraphics[width=\textwidth]{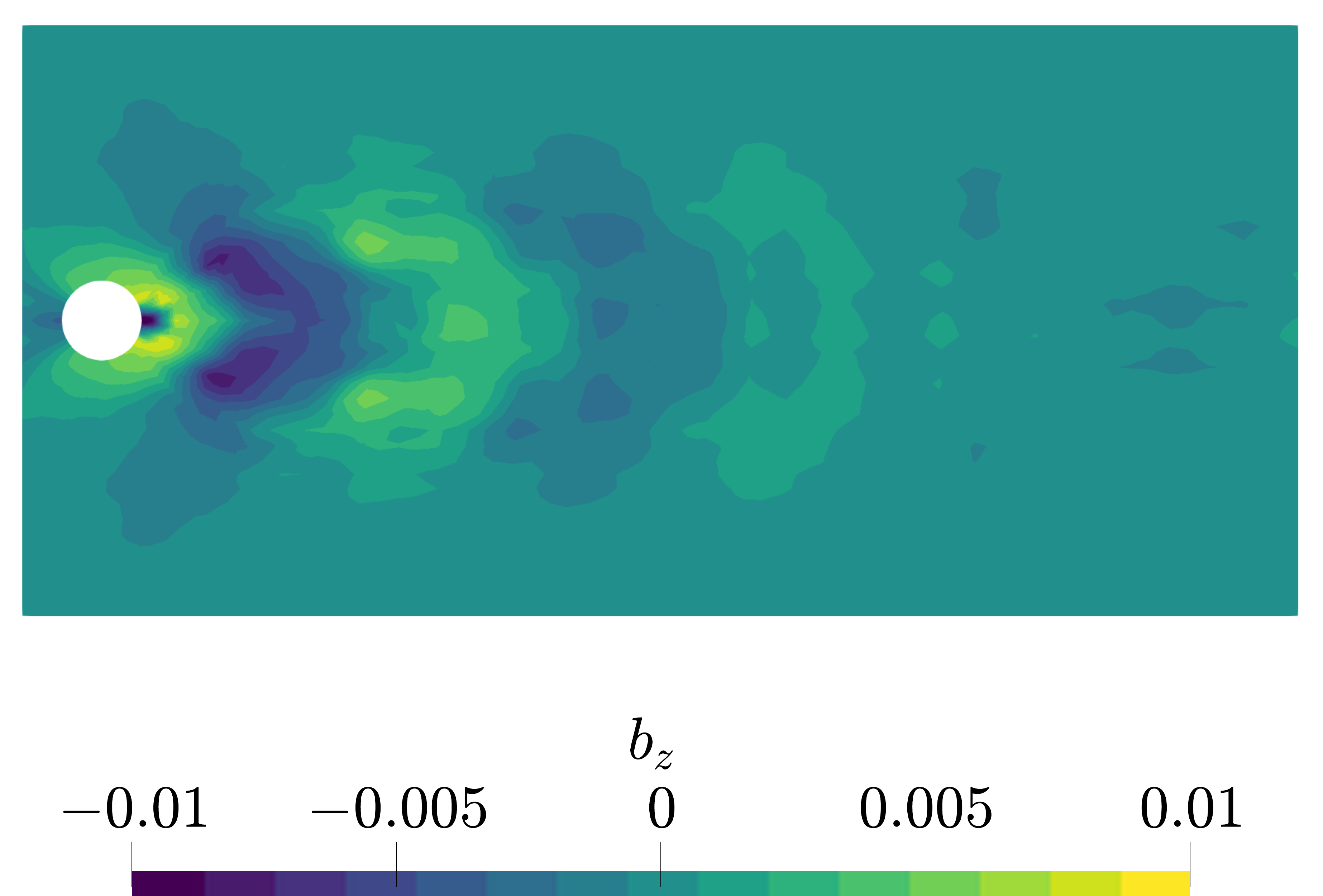}
     \end{subfigure}
     \hfill
     \begin{subfigure}[t]{0.32\textwidth}
         \centering
         \includegraphics[width=\textwidth]{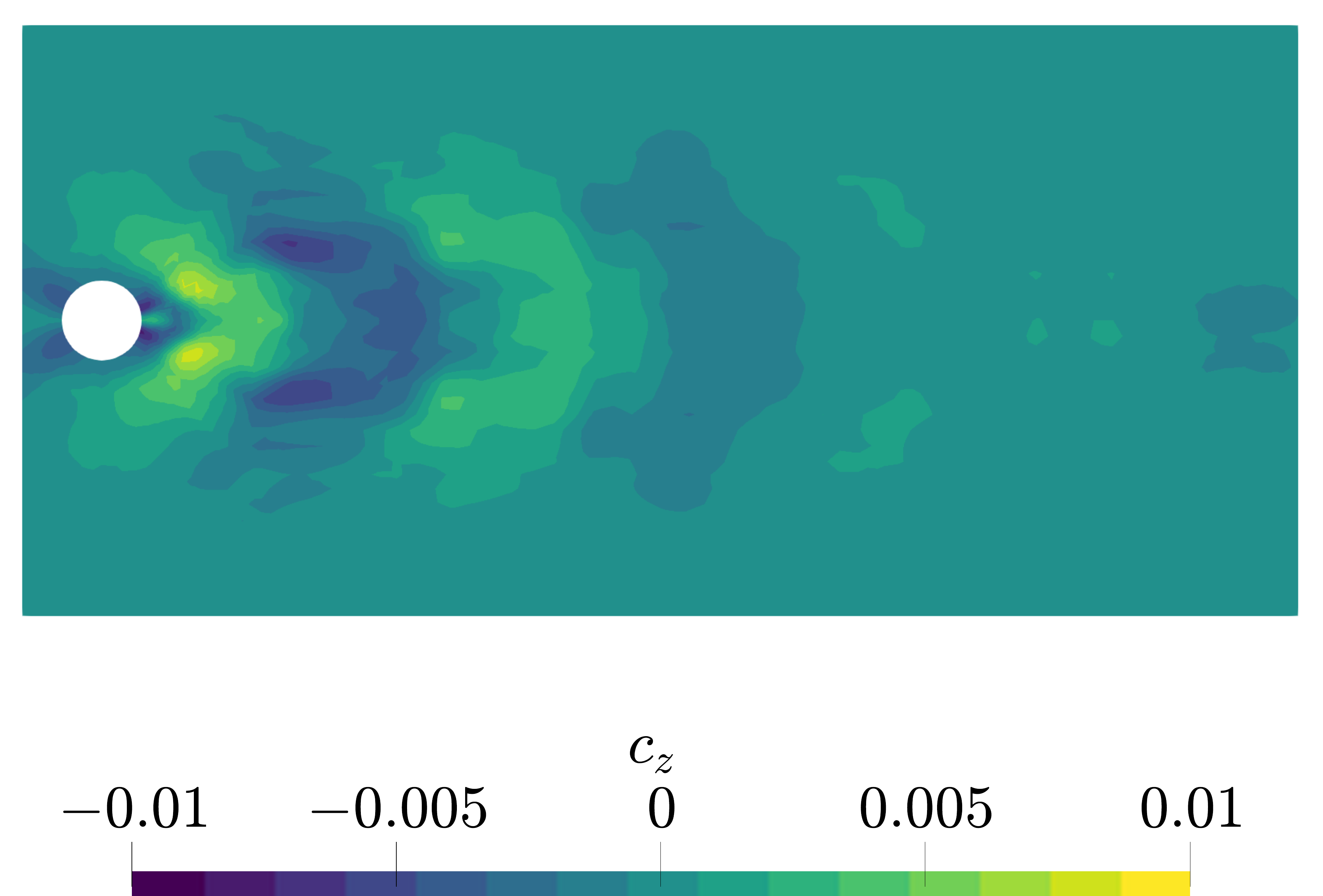}
     \end{subfigure}
\caption{Stationary parameters after optimization of zeroth and first Fourier modes. For better visualization purposes the parameter fields are shown for $-D \leq x \leq 15D$ and $-3.7D \leq y \leq 3.7 D$.}
\label{fig:data assimilation parameters}
\end{figure}

The stationary data assimilation parameters obtained from zeroth and first Fourier mode optimization are depicted in fig.~\ref{fig:data assimilation parameters}. Since a two-dimensional flow is studied, only the $z$-components of the data assimilation parameters are tuned. Although the parameter field $a_z$ (zeroth mode) is very smooth, the parameter fields $b_z$ (real component of the first mode) and $c_z$ (imaginary component of the first mode) exhibit some noise. This is due to the regularization weight which is set higher for $a_z$ than for $b_z$ and $c_z$. However, the streamwise velocity profiles obtained for the first Fourier mode (cf.~figs.~\ref{fig:velocity profiles circular cylinder wake mode 1, re after 0} and \ref{fig:velocity profiles circular cylinder wake mode 1, im after 0}) still are smooth. As discussed in previous works~\cite{brenner22,brenner23,plogmann23}, the higher the regularization weight is chosen, the smoother the parameter field and hence the velocity profiles become. Though, the optimization is more restricted, since possible changes in the parameter field are penalized by a contribution to the cost function.

The biggest magnitude of the parameter fields is evidently found in the near-cylinder and near-wake regions and slowly decays downstream and towards the top and bottom boundaries.

\section{Conclusions and outlook}
\label{sec:conclusion}

A variational data assimilation approach for unsteady RANS simulations based on the stationary discrete adjoint method is proposed. For this, the URANS equations are solved in physical space and time, and a discrete-time Fourier transform is performed assuming a time-periodic flow. Therefore, the corresponding discrete adjoint equations are stationary, and data is assimilated in the Fourier domain. As data assimilation parameter, an unsteady divergence-free source term is added to the URANS equations to correct for closure model discrepancies. Since this unsteady parameter is expressed by a Fourier series expansion, it is represented by the (stationary) Fourier coefficients in the Fourier domain with respect to the Fourier mode under investigation. Even though time-resolved data still is needed to obtain Fourier transformed reference data, time-stepping schemes as needed in a dynamic data assimilation framework (e.\,g.~4DVar) do not need to be employed. Making use of our efficient semi-analytical approach to compute the cost function gradient and the gradient-based \textit{Demon Adam} optimizer for the parameter update, we demonstrated that for coarse URANS simulations, sparse Fourier transformed (velocity) reference data can be assimilated through a 3DVar approach, avoiding the use of the very costly 4DVar method. Furthermore, a standard fluid solver (here based on a finite volume method implemented in \textit{OpenFOAM}) is exploited, and the use of a harmonic solver is avoided, making this method an attractive tool for more complex flows. 

The efficacy of the approach was established for flow around a two-dimensional circular cylinder at a Reynolds number of 3900. Sparse reference data was synthetically generated by means of a source term in the URANS momentum equations. To highlight the basic capabilities of the novel method, data assimilation has been performed for the zeroth and first Fourier modes. In particular, the assimilation of zeroth mode data allows for mean flow reconstruction as well as an improved vortex shedding frequency (Strouhal number), while the first mode data assimilation improved the low-frequency dynamics of the present flow.

Future work will focus on applying the proposed method to higher Fourier modes as well as evaluating the robustness of this framework with respect to reference data from high-fidelity experiments and simulations. Moreover, the applicability of the method to other flow setups at different Reynolds numbers should be examined. This could be combined with a more in-depth analysis of flows that do not exhibit a perfectly periodic behavior in time. 

We would also like to point out that this method is a very promising candidate for data assimilation in the context of large eddy simulations to improve their predictive capabilities.


\section*{Acknowledgments}
\label{sec:acknowledgments}

The authors would like to thank Arthur Couteau for fruitful discussions.


\appendix

\setcounter{figure}{0}    

\section{Fourier transform}
\label{app:Fourier transform}

\subsection{Zeroth Fourier mode}
\label{app:zeroth fourier mode}

In order to construct the Fourier transformed adjoint equation~\eqref{eq:fourier_urans_momentum_0} for the zeroth mode, every term from eq.~\eqref{eq:rans_momentum} is Fourier transformed according to eq.~\eqref{eq:fourier_transform} with $k=0$, i.\,e., $\omega^0 =0$. Recalling the assumption of a time-periodic flow signal, the time-derivative term reads
\begin{equation}
\frac{1}{N} \sum_{l=0}^{N-1} \frac{\partial \bar{u}_{i}(t_l)}{\partial t} e^{-i \omega^{0} t_l} = 0
\, ,
\end{equation}
while for the convection term it follows
\begin{equation}
\frac{1}{N} \sum_{l=0}^{N-1} \frac{\partial \bar{u}_{i}(t_l) \bar{u}_{j}(t_l)}{\partial x_{j}} e^{-i \omega^{0} t_l} = \frac{\partial \hat{u}_{i}^0 \hat{u}_{j}^0}{\partial x_{j}} + \frac{\partial \widehat{\bar{u}_l''\bar{u}_j''}^0}{\partial x_j}
\, .
\end{equation}

The pressure gradient is transformed as
\begin{equation}
\frac{1}{N} \sum_{l=0}^{N-1} \frac{\partial \bar{p}(t_l)}{\partial x_{i}}  e^{-i \omega^{0} t_l} = \frac{\partial \hat{p}^0}{\partial x_{i}} 
\end{equation}
and the Fourier transformed diffusion term yields
\begin{equation}
\frac{1}{N} \sum_{l=0}^{N-1} \frac{\partial}{\partial x_{j}}
    \left[
        \left(\nu + \nu_t\right)
        \left(\frac{\partial \bar{u}_i(t_l)}{\partial x_{j}} + \frac{\partial \bar{u}_j(t_l)}{\partial x_{i}}\right)
    \right]  e^{-i \omega^{0} t_l} = 
    \frac{\partial}{\partial x_{j}} 
    \left[
        \left(
        \nu + \hat{\nu}^0_t
        \right)
        \left(
        \frac{\partial \hat{u}_i^0}{\partial x_{j}} + \frac{\partial \hat{u}_j^0}{\partial x_{i}}
        \right)
    \right]
    +
    \frac{\partial}{\partial x_j} \left( 2 \widehat{\nu_t'' \bar{S}_{jl}''}^0 \right)
    \, .
\end{equation}

Fourier transforming the time-dependent forcing (see eq.~\eqref{eq:data_assimilation_firstmode}) reads
\begin{equation}
\frac{1}{N} \sum_{l=0}^{N-1} \epsilon_{ijk} \frac{\partial \psi_k(t_l)}{\partial x_{j}}  e^{-i \omega^{0} t_l} = \epsilon_{ijk} \frac{\partial a_k}{\partial x_{j}}
\, .
\end{equation}

\subsection{First Fourier mode}
\label{app:first fourier mode}

Further, to construct the Fourier transformed adjoint equation~\eqref{eq:fourier_urans_momentum_1} for the first mode, every term from eq.~\eqref{eq:rans_momentum} is Fourier transformed according to eq.~\eqref{eq:fourier_transform} with $k=1$. The time-derivative term reads
\begin{equation}
\frac{1}{N} \sum_{l=0}^{N-1} \frac{\partial \bar{u}_{i}(t_l)}{\partial t} e^{-i \omega^{1} t_l} = i \omega^1 \hat{u}_i^1
\, ,
\end{equation}
while for the convection term it follows
\begin{equation}
\frac{1}{N} \sum_{l=0}^{N-1} \frac{\partial \bar{u}_{i}(t_l) \bar{u}_{j}(t_l)}{\partial x_{j}} e^{-i \omega^{1} t_l} = \frac{\partial}{\partial x_{j}} \left( \hat{u}_{i}^0 \hat{u}_{j}^1 + \hat{u}_{i}^1 \hat{u}_{j}^0 \right) + \frac{\partial \widehat{\bar{u}_l''\bar{u}_j''}^1}{\partial x_j}
\, .
\end{equation}

The pressure gradient is transformed as
\begin{equation}
\frac{1}{N} \sum_{l=0}^{N-1} \frac{\partial \bar{p}(t_l)}{\partial x_{i}}  e^{-i \omega^{1} t_l} = \frac{\partial \hat{p}^1}{\partial x_{i}} 
\end{equation}
and the Fourier transformed diffusion term yields
\begin{equation}
\begin{split}
    \frac{1}{N} \sum_{l=0}^{N-1} \frac{\partial}{\partial x_{j}}
    \left[
        \left(\nu + \nu_t\right)
        \left(\frac{\partial \bar{u}_i(t_l)}{\partial x_{j}} + \frac{\partial \bar{u}_j(t_l)}{\partial x_{i}}\right)
    \right]  e^{-i \omega^{1} t_l}
    = 
    &\frac{\partial}{\partial x_{j}} 
    \left[
        \left(
        \nu + \hat{\nu}^0_t
        \right)
        \left(
        \frac{\partial \hat{u}_i^1}{\partial x_{j}} + \frac{\partial \hat{u}_j^1}{\partial x_{i}}
        \right)
        +
        \hat{\nu}^1_t
        \left(
        \frac{\partial \hat{u}_i^0}{\partial x_{j}} + \frac{\partial \hat{u}_j^0}{\partial x_{i}}
        \right)
    \right] \\
    &+
    \frac{\partial}{\partial x_j} \left( 2 \widehat{\nu_t'' \bar{S}_{jl}''}^1 \right)
    \, .
\end{split}
\end{equation}

Fourier transforming the time-dependent forcing (see eq.~\eqref{eq:data_assimilation_firstmode}) reads
\begin{equation}
\frac{1}{N} \sum_{l=0}^{N-1} \epsilon_{ijk} \frac{\partial \psi_k(t_l)}{\partial x_{j}}  e^{-i \omega^{1} t_l} = \epsilon_{ijk} \frac{\partial}{\partial x_{j}} \left( \frac{b_k}{2} - \frac{c_k}{2}i\right)
\, .
\end{equation}

\section{Adjoint equation}
\label{app: sequential optimization for k modes}

As discussed in section~\ref{sec:Sequential optimization procedure for zeroth and first Fourier mode}, the optimization is performed sequentially in terms of the Fourier modes. In conjunction with the definition of the cost function in eq.~\eqref{eq:cost_function}, due to the use of weights $W^k$, the coupling terms from the zeroth to the first Fourier mode (cf.~eq.~\eqref{eq:lagrangian_multiplier_coupled}) are made redundant. This implicates that an extension to the assimilation of $k$ Fourier modes seems straight forward. In that case, the decoupled adjoint equation reads
\begin{equation}
\label{sec:adjoint_problem_k_modes}
\begingroup
\renewcommand*{\arraystretch}{1.3}
\begin{bmatrix}
\mathbf{A}^0 & 0 & \cdots & 0 \\
0 & \mathbf{A}^1 & \cdots & 0 \\
\vdots & \vdots & \ddots & \vdots \\
0 & 0 & \cdots & \mathbf{A}^{N - 1}
\end{bmatrix}
\begin{bmatrix}
\lambda^0 \\
\lambda^1 \\
\vdots \\
\lambda^{N-1} \\
\end{bmatrix}
=
\begin{bmatrix}
\left(\frac{\partial f_{\hat{U}}}{\partial \hat{U}^0}\right)^T \\
\left(\frac{\partial f_{\hat{U}}}{\partial \hat{U}^1}\right)^T \\
\vdots \\
\left(\frac{\partial f_{\hat{U}}}{\partial \hat{U}^{N-1}}\right)^T \\
\end{bmatrix}
\endgroup
\, ,
\end{equation}
where the coupling terms are omitted allowing a sequential solution of $\lambda^k$ and hence a sequential optimization for each mode, starting from the zeroth mode up to the mode of truncation.

This extension, however, still is to be properly tested and subject of future investigations.

\section{Adjoint gradient}
\label{app:adjoint gradient}

The adjoint gradient with respect to parameters $a$, $b$ and $c$ is computed as
\begin{equation}
\label{eq:adjoint_gradient_derivation}
\begin{split}
\frac{\mathrm{d}f}{\mathrm{d}\hat{\psi}}
=
\begin{bmatrix}
\frac{\mathrm{d}f}{\mathrm{d}a} \quad \frac{\mathrm{d}f}{\mathrm{d}b} \quad \frac{\mathrm{d}f}{\mathrm{d}c}
\end{bmatrix}
&=
\begin{bmatrix}
\frac{\partial f}{\partial a} & \frac{\partial f}{\partial b} & \frac{\partial f}{\partial c}
\end{bmatrix}
- \lambda^T
\begin{bmatrix}
\frac{\partial \hat{R}}{\partial a} & \frac{\partial \hat{R}}{\partial b} & \frac{\partial \hat{R}}{\partial c}
\end{bmatrix} \\
&=
\begin{bmatrix}
\frac{\partial f}{\partial a} & \frac{\partial f}{\partial b} & \frac{\partial f}{\partial c}
\end{bmatrix}
-
\begin{bmatrix}
\left(\lambda_{\mathrm{R}}^{0}\right)^T & \left(\lambda_{\mathrm{I}}^{0}\right)^T & \left(\lambda_{\mathrm{R}}^{1}\right)^T & \left(\lambda_{\mathrm{I}}^{1}\right)^T
\end{bmatrix}
\begin{bmatrix}
\frac{\partial \hat{R}_{\mathrm{R}}^0}{\partial a} & \frac{\partial \hat{R}_{\mathrm{R}}^0}{\partial b} & \frac{\partial \hat{R}_{\mathrm{R}}^0}{\partial c} \\
\frac{\partial \hat{R}_{\mathrm{I}}^0}{\partial a} & \frac{\partial \hat{R}_{\mathrm{I}}^0}{\partial b} & \frac{\partial \hat{R}_{\mathrm{I}}^0}{\partial c} \\
\frac{\partial \hat{R}_{\mathrm{R}}^1}{\partial a} & \frac{\partial \hat{R}_{\mathrm{R}}^1}{\partial b} & \frac{\partial \hat{R}_{\mathrm{R}}^1}{\partial c} \\
\frac{\partial \hat{R}_{\mathrm{I}}^1}{\partial a} & \frac{\partial \hat{R}_{\mathrm{I}}^1}{\partial b} & \frac{\partial \hat{R}_{\mathrm{I}}^1}{\partial c} \\
\end{bmatrix}\\
&=
\begin{bmatrix}
\frac{\partial f}{\partial a} & \frac{\partial f}{\partial b} & \frac{\partial f}{\partial c}
\end{bmatrix}
-
\begin{bmatrix}
\left(\lambda_{\mathrm{R}}^{0}\right)^T \frac{\partial \hat{R}_{\mathrm{R}}^0}{\partial a} & 
\left(\lambda_{\mathrm{R}}^{1}\right)^T \frac{\partial \hat{R}_{\mathrm{R}}^1}{\partial b} & \left(\lambda_{\mathrm{I}}^{1}\right)^T \frac{\partial \hat{R}_{\mathrm{I}}^1}{\partial c}
\end{bmatrix}
\, .
\end{split}
\end{equation}

The derivative of the forward problem residual with respect to the parameter $a$ reads
\begin{equation}
    \label{eq:dRdPsi_a}
    \frac{\partial \hat{R}_{\mathrm{R}}^0}{\partial a}
    =
    \frac{\partial}{\partial a}
    \left[
        -\nabla\times a
    \right]
    \approx
    -\frac{\partial}{\partial a}
    \left[
        \mathbf{A}_{\hat{\psi}} a - b_{a}
    \right]
    =
    -\mathbf{A}_{\hat{\psi}}
    \, ,
\end{equation}
while the derivative of the forward problem residual with respect to the parameter $b$ yields
\begin{equation}
    \label{eq:dRdPsi_b}
    \frac{\partial \hat{R}_{\mathrm{R}}^1}{\partial b}
    =
    \frac{\partial}{\partial b}
    \left[
        -\nabla\times\left(\frac{b}{2}\right)
    \right]
    \approx
    -\frac{\partial}{\partial b}
    \left[
        \frac{1}{2}\mathbf{A}_{\hat{\psi}} b - b_{b}
    \right]
    =
    -\frac{1}{2} \mathbf{A}_{\hat{\psi}}
    \, ,
\end{equation}
and lastly, with respect to the parameter $c$ is defined as
\begin{equation}
    \label{eq:dRdPsi_c}
    \frac{\partial \hat{R}_{\mathrm{I}}^1}{\partial c}
    =
    \frac{\partial}{\partial c}
    \left[
        \nabla\times\left(\frac{c}{2}\right)
    \right]
    \approx
    \frac{\partial}{\partial c}
    \left[
        \frac{1}{2}\mathbf{A}_{\hat{\psi}} c - b_{c}
    \right]
    =
    \frac{1}{2}\mathbf{A}_{\hat{\psi}}
    \, .
\end{equation}

The derivatives thus correspond to the system matrix $\mathbf{A}_{\hat{\psi}}$. Combining these approximate terms with eq.~\eqref{eq:adjoint_gradient_derivation} yields
\begin{equation}
\frac{\mathrm{d}f}{\mathrm{d}\hat{\psi}}
=
\begin{bmatrix}
\frac{\mathrm{d}f}{\mathrm{d}a} \quad \frac{\mathrm{d}f}{\mathrm{d}b} \quad \frac{\mathrm{d}f}{\mathrm{d}c}
\end{bmatrix}
=
    \begin{bmatrix}
    \left(\frac{\partial f}{\partial a} + \left(\lambda_{\mathrm{R}}^{0}\right)^T \mathbf{A}_{\hat{\psi}}\right) & \left(\frac{\partial f}{\partial b} + \frac{1}{2}\left(\lambda_{\mathrm{R}}^{1}\right)^T \mathbf{A}_{\hat{\psi}}\right)  &
    \left(\frac{\partial f}{\partial c} - \frac{1}{2}\left(\lambda_{\mathrm{I}}^{1}\right)^T \mathbf{A}_{\hat{\psi}}\right) 
    \end{bmatrix}
\end{equation}
for the final expression of the adjoint gradient with respect to all three stationary parameters.

\section{Implementation}
\label{app:Implementation}

The computational mesh was created using \textit{gmsh}~\cite{geuzaine09}, and the cell size is decreasing toward the cylinder wall to capture the flow detachment well enough. However, the wall-normal distance still is quite large to conform with the requirements of wall functions. This makes the mesh relatively coarse and allows for larger time steps, which in turn decreases the CPU time of the solution of the forward problem drastically.

Second order schemes are used for discretization, and a bi-conjugate gradient stabilized linear solver with a Cholesky preconditioner is used for the forward coupled system, whereas no preconditioning is applied to the adjoint coupled system. The implicit Euler scheme is chosen for time integration, while a constant time step is used for uniform sampling as part of the spectral analysis (Fourier transform). The time step size ensures an average Courant-Friedrichs-Lewy (CFL) number smaller than 0.3. For this particular work the two-equation $k$-$\omega$ SST model as proposed by \cite{Menter2003} is used due to its ability to account for the transport of the principal shear stress in adverse pressure gradient boundary layers~\cite{menterReviewShearstressTransport2009}. Moreover, \textit{OpenFOAM} incorporates parallel computing as an inherent capability, utilizing domain decomposition. This involves dividing the geometry and its corresponding fields into segments, which are then assigned to individual processors. The parallel calculations are executed using the widely adopted \textit{Open MPI} implementation of the message passing interface (MPI) standard. Parallel computing is only used for the solution of the forward problem, but not for the adjoint problem solution.


\end{document}